\documentclass[sigconf,screen, 10pt]{acmart}
\usepackage{subcaption}

\usepackage{url}
\usepackage{enumitem}
\usepackage{graphicx}
\usepackage{caption}
\usepackage{amsmath}
\usepackage{tabu}
\usepackage{multirow}
\usepackage{hyperref}
\usepackage{adjustbox}
\usepackage{mathptmx}
\usepackage{amsfonts}
\usepackage{algpseudocode}
\usepackage{pgfplots}
\pgfplotsset{compat=1.12}
\usepackage{soul}
\usepackage{footnote}
\usepackage{lipsum}
\usepackage[ruled,vlined, linesnumbered]{algorithm2e}
\usepackage{subcaption}
\usepackage{listings}
\colorlet{punct}{red!60!black}
\definecolor{background}{HTML}{FFFFFF}
\definecolor{delim}{RGB}{20,105,176}
\colorlet{numb}{magenta!60!black}

\newcommand*\circled[1]{\tikz[baseline=(char.base)]{
            \node[shape=circle,draw,inner sep=0.75pt, text=white,fill=black] (char) {#1};}}
\lstdefinelanguage{json}{
    basicstyle=\footnotesize\ttfamily,
    numbers=left,
    numberstyle=\scriptsize,
    xleftmargin=2.3em,
    xrightmargin=0.5em,
    framexleftmargin=1.9em,
    stepnumber=1,
    numbersep=8pt,
    showstringspaces=false,
    breaklines=true,
    frame=single,
    backgroundcolor=\color{background},
    literate=
     *{0}{{{\color{numb}0}}}{1}
      {1}{{{\color{numb}1}}}{1}
      {2}{{{\color{numb}2}}}{1}
      {3}{{{\color{numb}3}}}{1}
      {4}{{{\color{numb}4}}}{1}
      {5}{{{\color{numb}5}}}{1}
      {6}{{{\color{numb}6}}}{1}
      {7}{{{\color{numb}7}}}{1}
      {8}{{{\color{numb}8}}}{1}
      {9}{{{\color{numb}9}}}{1}
      {:}{{{\color{punct}{:}}}}{1}
      {,}{{{\color{punct}{,}}}}{1}
      {\{}{{{\color{delim}{\{}}}}{1}
      {\}}{{{\color{delim}{\}}}}}{1}
      {[}{{{\color{delim}{[}}}}{1}
      {]}{{{\color{delim}{]}}}}{1},
}

\AtBeginDocument{%
  \providecommand\BibTeX{{%
    \normalfont B\kern-0.5em{\scshape i\kern-0.25em b}\kern-0.8em\TeX}}}

\acmYear{2021}\copyrightyear{2021}
\setcopyright{acmcopyright}
\acmConference[WoSC '21]{Workshop on Serverless Computing}{December 6--10, 2021}{Qu\'ebec, Canada}
\acmBooktitle{Workshop on Serverless Computing (WoSC '21), December 6--10, 2021, Qu\'ebec, Canada}
\acmPrice{15.00}
\acmISBN{978-1-4503-XXXX-X/20/12}

\begin{document}

\title{Towards Demystifying Intra-Function Parallelism in Serverless Computing}

\author{Michael Kiener}
\email{michael.kiener@tum.de}
\affiliation{%
 \institution{Chair of Computer Architecture and Parallel Systems \\ Technische Universit{\"a}t M{\"u}nchen}
  \city{Garching (near Munich)}
  \state{Germany}
}

\author{Mohak Chadha}
\email{mohak.chadha@tum.de}
\affiliation{%
 \institution{ Chair of Computer Architecture and Parallel Systems \\ Technische Universit{\"a}t M{\"u}nchen}
  \city{Garching (near Munich)}
  \state{Germany}
}

\author{Michael Gerndt}
\email{gerndt@in.tum.de}
\affiliation{%
     \institution{Chair of Computer Architecture and Parallel Systems \\ Technische Universit{\"a}t M{\"u}nchen}
  \city{Garching (near Munich)}
  \state{Germany}
}

\renewcommand{\shortauthors}{Michael Kiener et al.}

\begin{abstract}

Serverless computing offers a pay-per-use model with high elasticity and automatic scaling for a wide range of applications. Since cloud providers abstract most of the underlying infrastructure, these services work similarly to black-boxes. As a result, users can influence the resources allocated to their functions, but might not be aware that they have to parallelize them to profit from the additionally allocated virtual CPUs (vCPUs). In this paper, we analyze the impact of parallelization within a single function and container instance for AWS Lambda, Google Cloud Functions (GCF), and Google Cloud Run (GCR). We focus on compute-intensive workloads since they benefit greatly from parallelization. Furthermore, we investigate the correlation between the number of allocated CPU cores and vCPUs in serverless environments. Our results show that the number of available cores to a function/container instance does not always equal the number of allocated vCPUs. By parallelizing serverless workloads, we observed cost savings up to 81\% for AWS Lambda, 49\% for GCF, and 69.8\% for GCR.

\end{abstract}



\maketitle

\section{Introduction}
\label{sec:intro}
With the advent of Amazon Web Services (AWS) Lambda in 2014, serverless computing has gained popularity and more adoption in different application domains such as machine learning~\cite{Chadha2020, carreira2019cirrus, Jiang2021}, scientific computing~\cite{nanopore, chard2020funcx, jindal2021function, postericdcs}, and linear algebra~\cite{serverlesslin}. In serverless computing, developers do not have to manage infrastructure themselves but completely hand over this responsibility to a Function-as-a-Service Platform. Several open-source and commercial FaaS platforms such as OpenWhisk, Google Cloud Functions (GCF), and Lambda are currently available. Applications are developed as small units of code, called functions that are independently packaged and uploaded to a FaaS platform and executed on event triggers such as HTTP requests. On function invocation, the FaaS platform creates an \emph{execution environment} (instance) which provides a secure and isolated runtime environment for the function. The functions can be written using various languages such as \texttt{Java}, \texttt{Go}, or \texttt{Nodejs} and a language-specific environment called as \emph{runtime} is created in the function's execution environment. However, due to the limitations on the available \emph{runtimes} in commercial FaaS platforms such as GCF, serverless Container-as-a-Service (CaaS) offerings such as Google Cloud Run (GCR)~\cite{GCR} have been introduced. GCR is a fully-managed service based on Knative~\cite{knative}. CaaS provides developers greater flexibility and allows them to build custom container images for their functions.

\begin{figure}[t]
\centering
\includegraphics[width=0.49\textwidth]{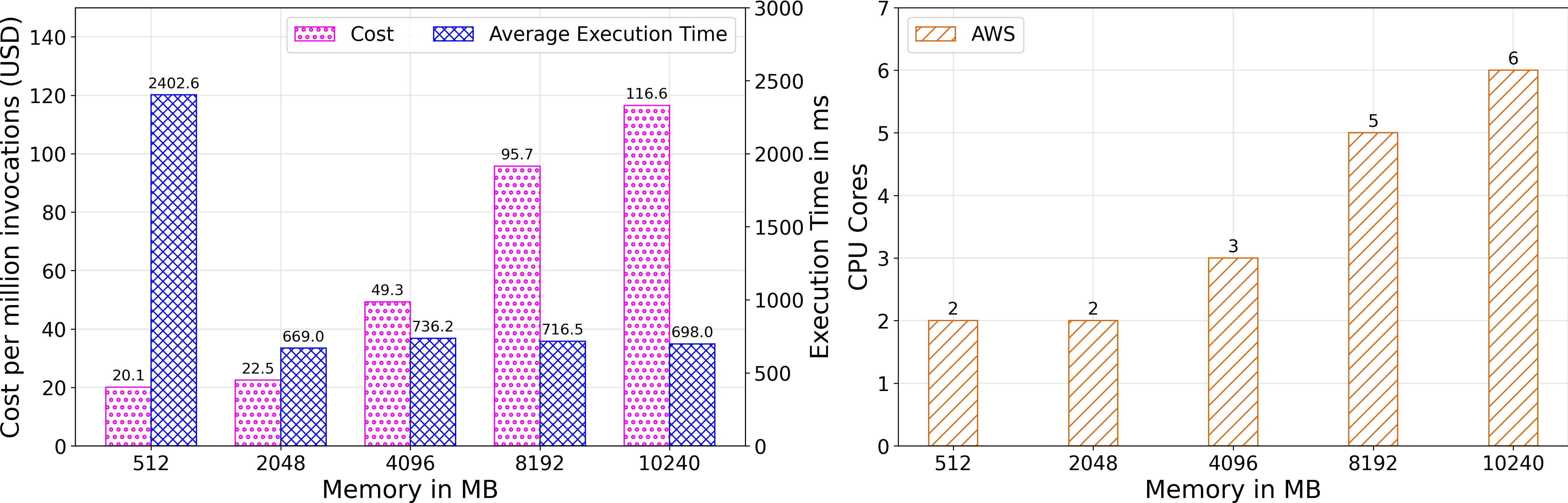}
\caption{Left: average execution time and cost for the modified \texttt{MVT} benchmark~\cite{npbench} in C++ using AWS Lambda (\texttt{us-east1}). Right: Number of CPU cores allocated for the different memory profiles for AWS Lambda.}
\label{fig:motivation_example}
\end{figure}

While most details about the backend infrastructure management are abstracted away from the user by commercial FaaS and serverless CaaS platforms, they still allow developers to configure the amount of memory and number of vCPUs (GCR) allocated to a function/container instance~\cite{LambdaConfig, GCRConfig}. As a result, each function/container instance has a fixed number of CPU cores and memory associated with it. For commercial FaaS platforms such as Lambda and GCF, the performance of the function is directly related to the amount of function memory configured. This is because these platforms increase the amount of compute capacity available to a function by increasing the number of vCPUs or the fraction of the CPU time if more memory is configured~\cite{behind}. Serverless is advertised as a pay-per-use model, where the users are billed based on the execution time of the functions measured in 100ms (GCR/GCF) and 1ms (Lambda) intervals. However, due to the billing policies followed by cloud providers, increasing the amount of memory often leads to an increase in costs due to fees payment wrt GB-Second (and GHz-Second with GCF/GCR~\cite{GCFPricing, GCRPricing}). The comparison between the average execution time and cost~\cite{LambdaPricing} for the modified \texttt{MVT} benchmark~\cite{npbench} when deployed on Lambda is shown in Figure~\ref{fig:motivation_example}. Although the average execution time decreases when more memory is configured, the cost increases significantly. Furthermore, after a certain memory configuration ($>$2048MB) allocating more memory does not considerably affect the function execution time. Serverless FaaS/CaaS platforms launch the function instances on the platform's traditional Infrastructure-as-a-Service (IaaS) virtual machines (VM) offerings~\cite{behind, chadha2021architecture}. However, the provisioning of such VM's is abstracted away from the user. As a result, the user is not aware of the details of the provisioned VMs such as the number of CPU cores and the virtual CPUs (vCPUs). Figure~\ref{fig:motivation_example} shows the number of CPU cores available to the function for the different memory profiles on AWS Lambda. We obtain the number of available cores using the Linux \texttt{proc} filesystem. Since the native implementation of the \texttt{MVT} benchmark is single-threaded it cannot utilize the underlying cores leading to resource under-utilization. To this end, parallelization of functions can lead to a significant reduction in execution time and thus reduced costs.



Our key contributions are:
\begin{itemize}
   \item \textbf{Identification of \#CPU cores and vCPU allocations}: We investigate the number of allocated CPU cores in FaaS/CaaS platforms and how they are mapped to the   allocated vCPUs.
    \item \textbf{Intra-function parallel workloads}: We modify and parallelize three different compute-intensive serverless workloads\footnote{We use the term serverless workload and function interchangeably.}, using C++, Java and Go. We execute these workloads on AWS Lambda, GCF, and GCR, and analyze their execution times.
    \item \textbf{Cost analysis}: We demonstrate the benefits of parallelizing functions and discuss conditions when parallelization can be beneficial.
\end{itemize}

The rest of the paper is structured as follow. \S\ref{sec:relatedWork} describes the previous work on inter-function parallelism and current strategies for performance optimization of FaaS functions. \S\ref{sec:Methodology} describes our methodology. In \S\ref{sec:expResults}, our experimental results are presented. Finally, \S\ref{sec:conclusion} concludes the paper and presents an outlook.

\section{Related Work}
\label{sec:relatedWork}
The majority of previous work on parallelizing applications via FaaS has focused on splitting the workload via separate function instances, i.e., inter-function parallelism~\cite{serverlesslin, parallelworkloads, 10.1145/3361525.3361535}. Shankar et al.~\cite{serverlesslin} showed that the elasticity provided by serverless computing can be used to efficiently execute linear algebra algorithms, which are inherently parallel. They implemented a system to split a linear algebra algorithm into tasks which are then executed by AWS Lambda functions. The data between function instances is shared via a persistent object-store. With their system, they achieved   performance within a factor of two, as compared to a server-centric Message Passing Interface (MPI)~\cite{mpi_standard} implementation. Pons et al.~\cite{parallelworkloads} analyzed the performance of fork/join workflows using existing services such as AWS Step Functions~\cite{awsstep}, Azure Durable Functions~\cite{azuredurable}, and OpenWhisk Composer~\cite{owcomposer}. They found that, while OpenWhisk Composer offers the best performance, all of the function orchestration solutions have significant overhead for executing parallel workloads. In~\cite{10.1145/3361525.3361535}, the authors present a system called Crucial that allows developers to program highly-concurrent stateful applications with serverless architectures. Crucial ports multi-threaded applications to a serverless environment, by leveraging a distributed shared memory layer. For coordinating functions they used shared objects. With Crucial, the authors achieved performance results similar to that of an equivalent Spark cluster. While distributing workloads across function invocations can be useful due to the high elasticity, results show that it can still lead to significant communication and synchronization overhead. In contrast, we focus on intra-function parallelism where we parallelize functions and execute them within a single function instance.

Evaluating the general performance of FaaS platforms and improving the performance of FaaS functions has also been actively researched~\cite{chadha2021architecture, 10.1145/3447545.3451173}. In~\cite{10.1145/3447545.3451173}, the authors present the Serverless Application Analytics Framework (SAAF), to improve observability on the performance of FaaS functions on commercial FaaS platforms. SAAF currently supports multiple FaaS platforms and several different programming languages. In our previous work~\cite{chadha2021architecture}, we examined the underlying processor architectures on GCF and demonstrated the usage of Numba~\cite{numba}, a Just-in-Time (JIT) compiler based on LLVM for optimizing and improving the performance of compute-intensive Python based FaaS functions. We showed that the optimization of FaaS functions can improve performance by $18.2$x and save costs by $76.8$\%. However, all of the previous approaches evaluate the performance of single-threaded FaaS functions. In contrast, we evaluate and analyze the performance of parallelized FaaS functions. Furthermore, we investigate different function configurations and evaluate their respective parallel efficiency. Moreover, we demonstrate significant cost savings for parallelized functions as compared to their sequential implementations.

\section{Methodology}
\label{sec:Methodology}

In this section, we first describe the different compute-intensive serverless workloads used in this work. 
Following this, we describe the different language runtimes we used for adapting and modifying the different workloads. Finally, we describe our benchmarking workflow.




\subsection{Serverless Workloads}
\label{sec:faas_wokloads}
For our experiments, we chose two microbenchmarks, i.e., \texttt{Atax} and \texttt{MVT} from 
NPBench~\cite{npbench, ziogas2021npbench} and one application, i.e., Monte Carlo from PyPerf~\cite{pyperf}. Both \texttt{Atax} and 
\texttt{MVT} take a JSON file as input describing the input matrix and vector sizes. \texttt{Atax} performs
a matrix-vector product, followed by multiplying the result again with the matrix. On the other hand, \texttt{MVT}
performs two matrix-vector products, followed by adding the results to different vectors. The Monte Carlo simulation
estimates the digits of $\pi$. It generates random numbers in a $1x1$ square and counts all points for which
the distance to the center is less than 1. The ratio of points is an estimate of $\frac{\pi}{4}$ which is 
used to estimate $\pi$. It takes a JSON file as input specifying the number of iterations for the simulation. 
We port the different workloads, initially written in Python to the different language runtimes used in 
this work (\S\ref{sec:language_runtimes}).


\subsection{Language Runtimes}
\label{sec:language_runtimes}
To evaluate the different services, i.e., AWS Lambda, GCF, and GCR wrt the performance of parallelized functions, 
we chose multiple programming languages, i.e., C++, Go, and Java. We chose C++ since it is widely used for scientific 
computing in high performance computing applications~\cite{nanopore}. However, none of the major commercial FaaS platforms
support C++ by default. For executing C++ functions on AWS Lambda, we use the custom C++ Lambda runtime~\cite{AWSLambdaRuntimeC++}
based on the Lambda Runtime API~\cite{AWSLambdaRuntimeAPI}. For GCF, it was not possible to run C++ functions since it 
does not support custom runtimes. On the other hand, for GCR we use a custom docker image and install the required dependencies
to build the C++ function.

We chose Go since it is widely used and supported by default on Lambda and GCF. Furthermore, it 
was designed with concurrency in mind which simplifies parallelization of functions. As our final language, we chose Java not only due to its popularity but also due to differences in its design as compared to the other two languages. In contrast to C++ and Go, Java applications are compiled to bytecode which is executed by the Java Virtual Machine (JVM). Similar to Go, Java is also available by default on both Lambda and GCF. For running Go, C++, and Java based functions on GCR, we used custom docker images with the required dependencies.The details about the different language runtimes, i.e., their versions, compiler, compiler flags, and the different GCR images
is shown in Table~\ref{tab:runtimeInfo}.

For parallelizing the different serverless workloads (\S\ref{sec:faas_wokloads}) using the different languages, we first identified suitable regions using profilers. Following this, we used additional libraries or language-specific features to parallelize those regions. For C++, we used OpenMP which is commonly used for shared memory programming. OpenMP allows developers to annotate code using pragmas which are automatically used by the compiler to generate multi-threaded code. In the case of Go, we made use of \emph{goroutines}, which are lightweight threads having their own stack. For Java, we utilized the \emph{ExecutorService} class which allows developers to create a thread pool and submit tasks to be executed by the threads. While OpenMP supports automatic division of work between threads, for Go and Java, we had to manually split the workload between the threads.

\begin{table}[t]
\caption{Runtime configurations. Includes parallelization technique, version, compiler and flags. For GCR, the container deployment image is also mentioned.}
	\centering
\begin{adjustbox}{width=8.5cm,center}
    \begin{tabu}{|c|c|c|c|c|}
	\tabucline{-}

    \textbf{Language} & \textbf{Parallelization} & \textbf{Version} & \textbf{Compiler \& Flags} & \textbf{GCR Image} \\\tabucline{-}
    C++ & OpenMP & AWS: C++11, GCR: C++17 & \texttt{g++, -O3} & debian:buster-slim \\ \tabucline{-}
    Go & goroutines & AWS,GCR: 1.16, GCF: 1.13 & \texttt{gc, GOOS=linux} & debian:buster-slim \\ \tabucline{-}
    Java & ExecutorService & Java 11 & \texttt{OpenJDK 11} & maven:3.8-jdk-11 \\ \tabucline{-}
  
\end{tabu}
\end{adjustbox}
\label{tab:runtimeInfo}
\end{table}

\begin{figure}[t]
    \centering
    \includegraphics[width=1\columnwidth]{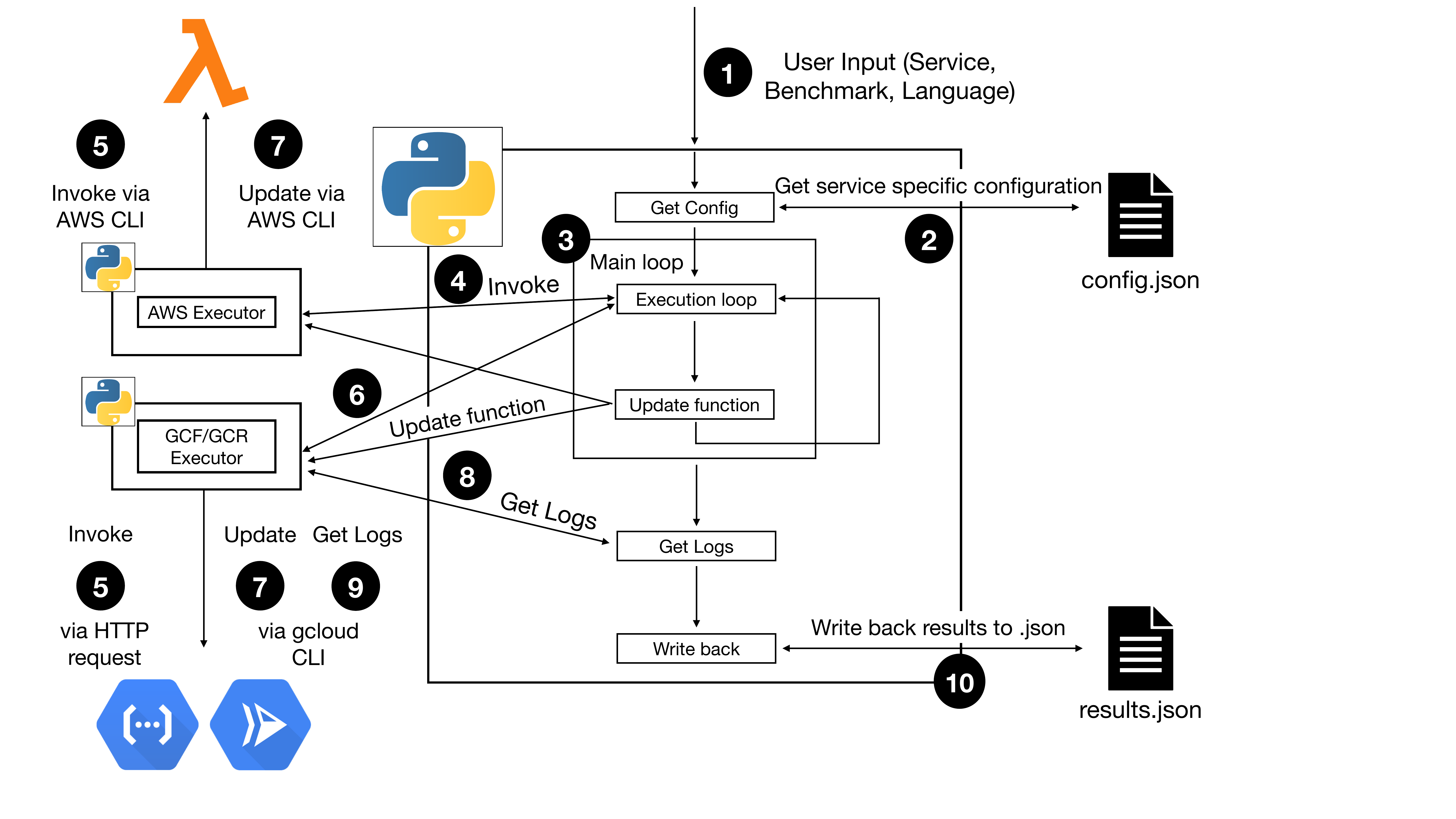}
    \caption{Different steps involved in our benchmarking workflow.}
    \label{fig:exec_script}
\end{figure}

\subsection{Benchmarking Workflow}

Initially, we deploy all the serverless wokloads (\S\ref{sec:faas_wokloads}) using the respective command line 
interfaces (CLIs) provided by AWS and Google. Figure~\ref{fig:exec_script} shows the different steps involved 
in our benchmarking process. To facilitate automatic function invocation and collection of function logs, we implement multiple Python scripts. At the beginning, the user provides the service type, serverless workload, and the language runtime (\S\ref{sec:language_runtimes}) as input to the Python script \circled{1}. Following this, the function configuration according to the input parameters is retrieved from a JSON file \circled{2}. The function configuration contains relevant parameters to invoke the function such as the function URL. In the main loop \circled{3} of the Python script, we repeatedly invoke the function synchronously, i.e., we await the function result before 
invoking it again \circled{4}. For Lambda, we use the \texttt{aws} CLI for invoking the functions, while for GCF/GCR the functions are triggered using HTTP requests \circled{5}. Each function takes a JSON file as input that specifies its input size and the number of threads to utilize. We execute each serverless workload 20 times, 10 times sequentially and 10 times with multiple threads. Following this, we update the memory configuration of the function \circled{6}. For this, we use their respective CLIs \circled{7}. After all function executions have finished, we store the results in a JSON file \circled{10}. For Lambda, all the required data can be retrieved from the function response, while for GCF/GCR we use the \texttt{gcloud} CLI \circled{8}, \circled{9}.

\section{Experimental Results}
\label{sec:expResults}
In this section, we describe our experimental setup and present results wrt performance and costs for the 
parallelized serverless workloads for the different services. Finally, we discuss the impact of cold starts in our experiments.

\begin{figure}[t]
\centering
\includegraphics[width=\columnwidth]{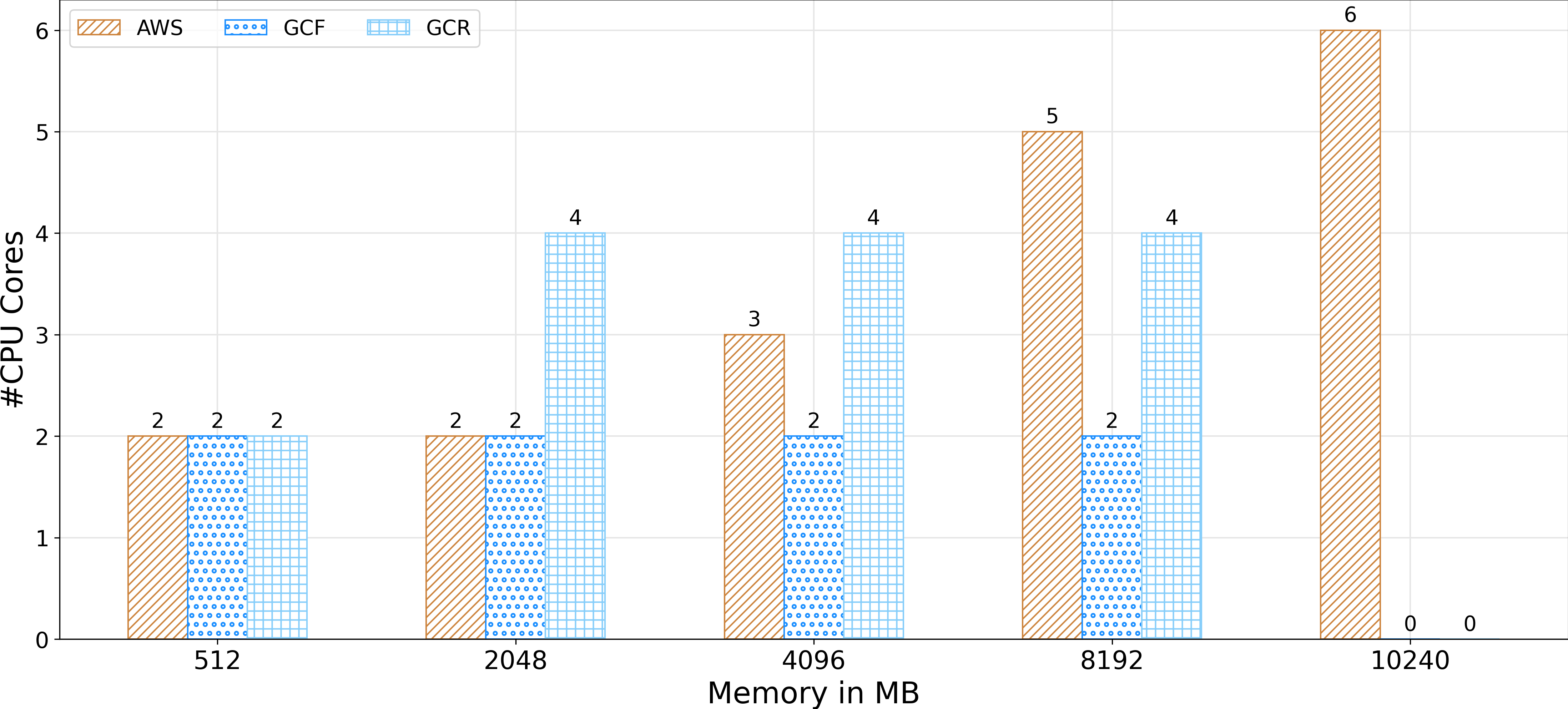}

\caption{Number of available CPU cores for the different services at the different memory configurations.}
\label{fig:numbercores}
\end{figure}

\begin{figure*}[t]
    \centering
    \begin{subfigure}{0.32\textwidth}
        \includegraphics[width=1\linewidth]{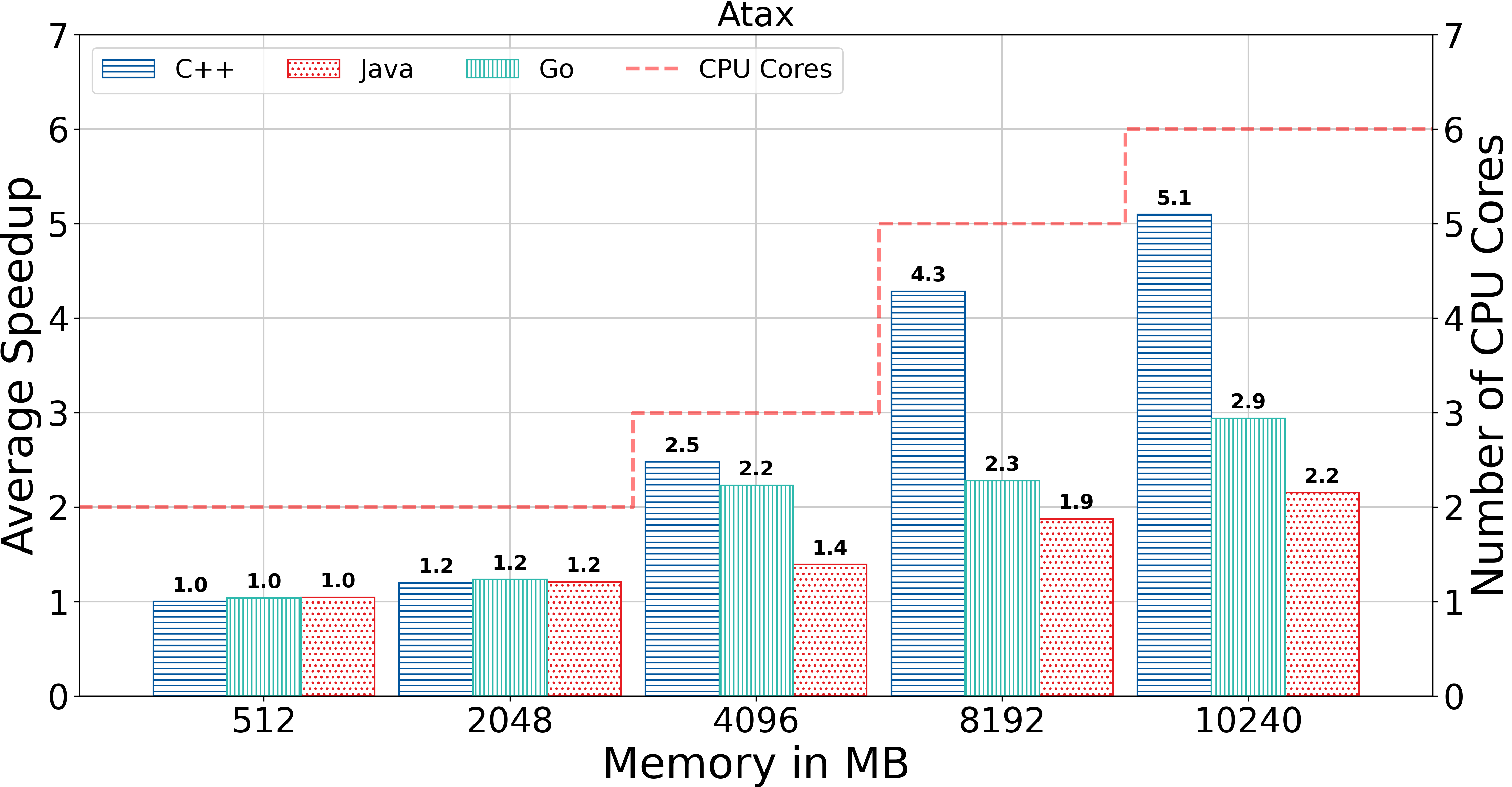}
        \caption{AWS Atax}
        \label{fig:speedups_aws_atax}
    \end{subfigure}
    \begin{subfigure}{0.32\textwidth}
        \includegraphics[width=1\linewidth]{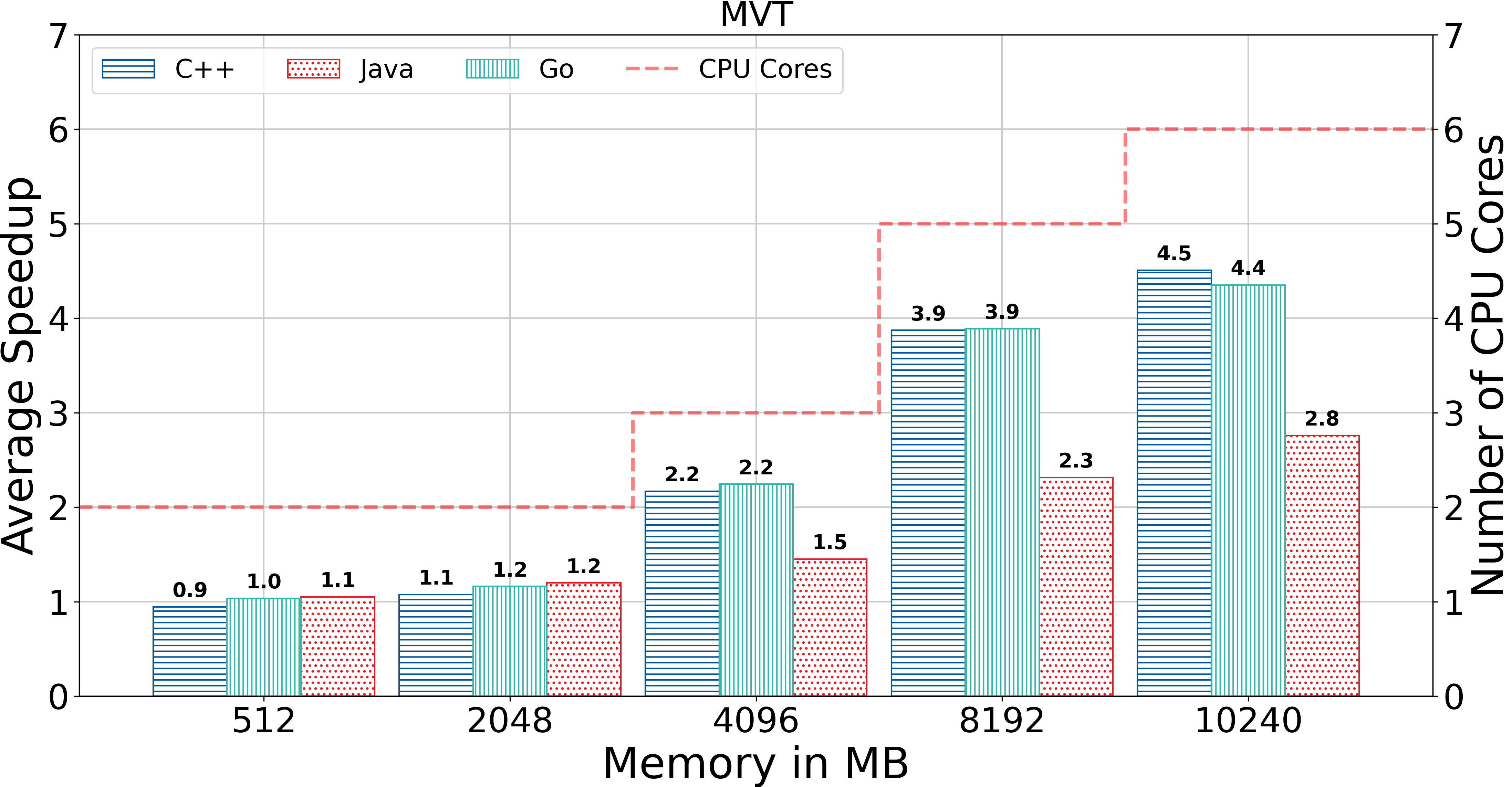}
        \caption{AWS MVT}
        \label{fig:speedups_aws_mvt}
    \end{subfigure}
    \begin{subfigure}{0.32\textwidth}
        \includegraphics[width=1\linewidth]{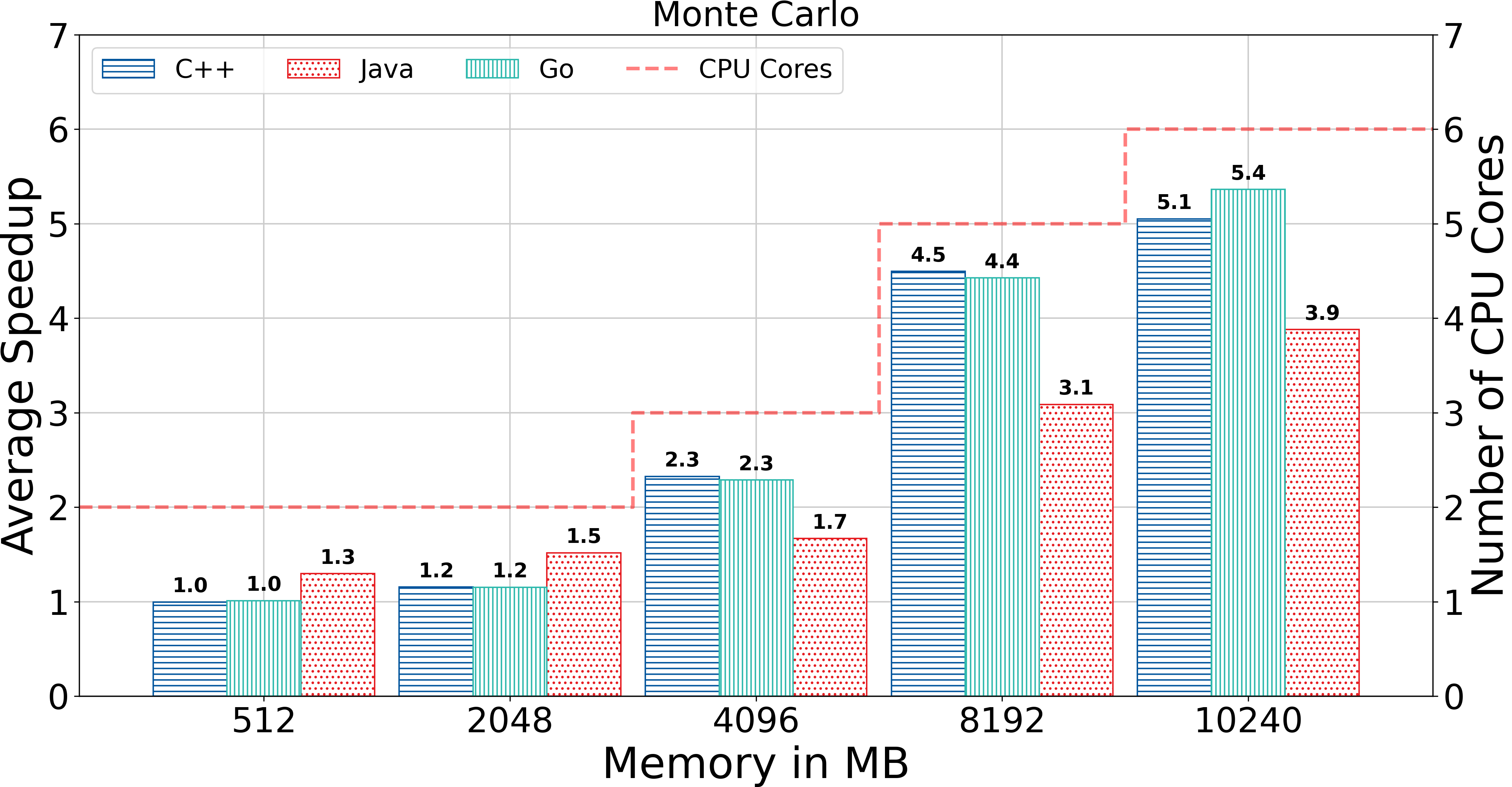}
        \caption{AWS Monte Carlo}
        \label{fig:speedups_aws_monte}
    \end{subfigure}
    \begin{subfigure}{0.32\textwidth}
        \includegraphics[width=1\linewidth]{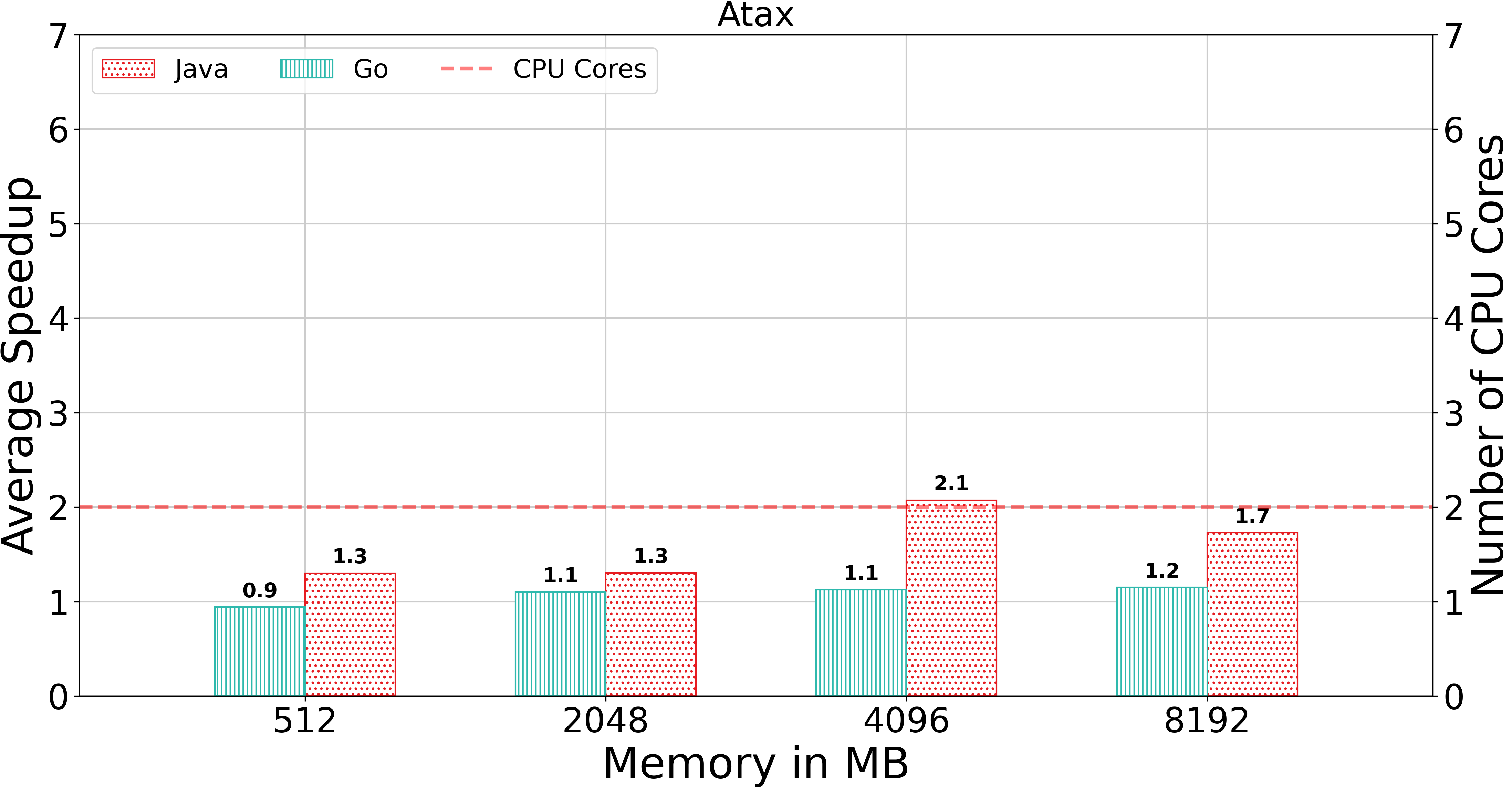}
        \caption{GCF Atax}
        \label{fig:speedups_gcf_atax}
    \end{subfigure}
    \begin{subfigure}{0.32\textwidth}
        \includegraphics[width=1\linewidth]{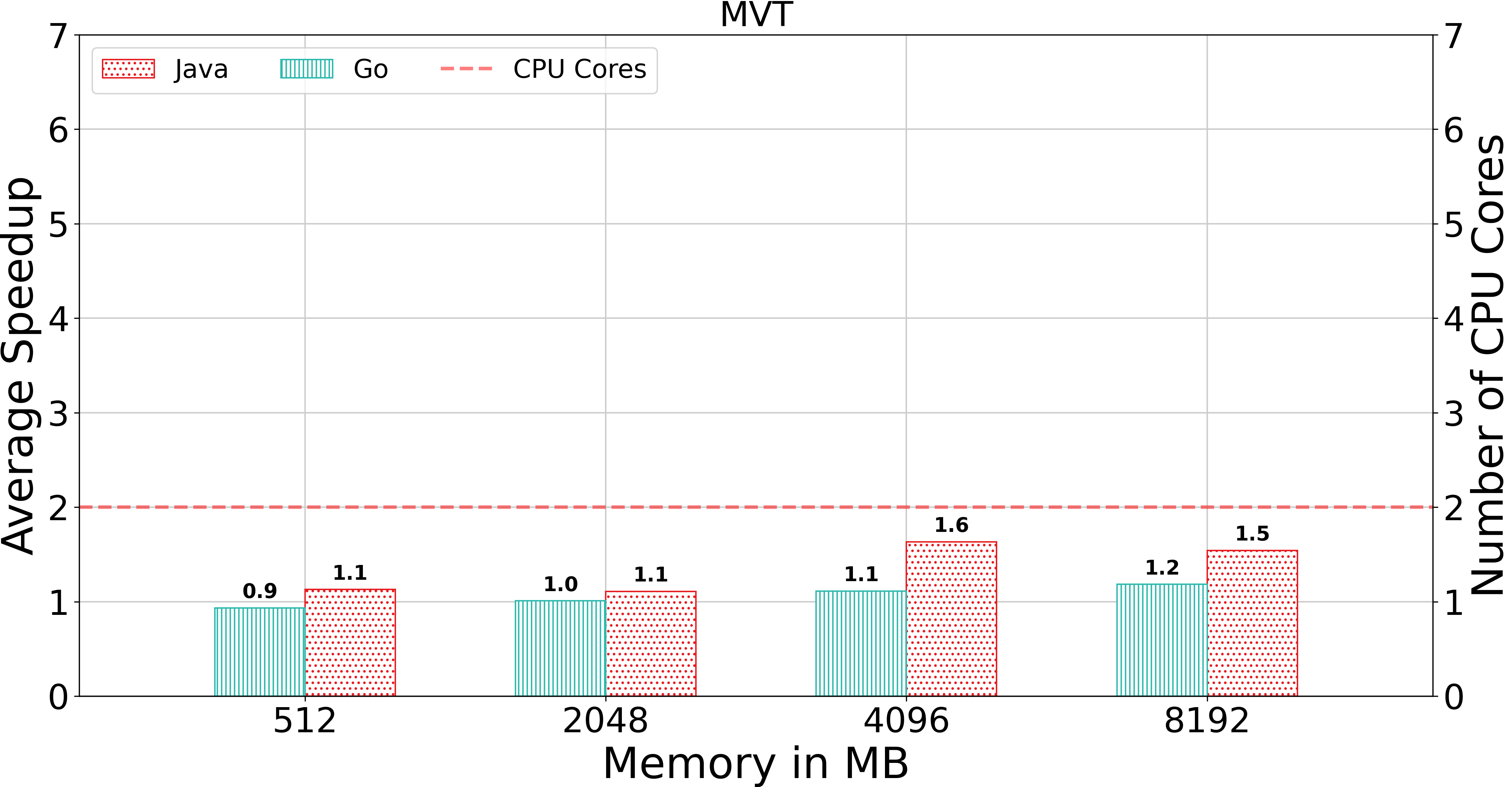}
        \caption{GCF MVT}
        \label{fig:speedups_gcf_mvt}
    \end{subfigure}
    \begin{subfigure}{0.32\textwidth}
        \includegraphics[width=1\linewidth]{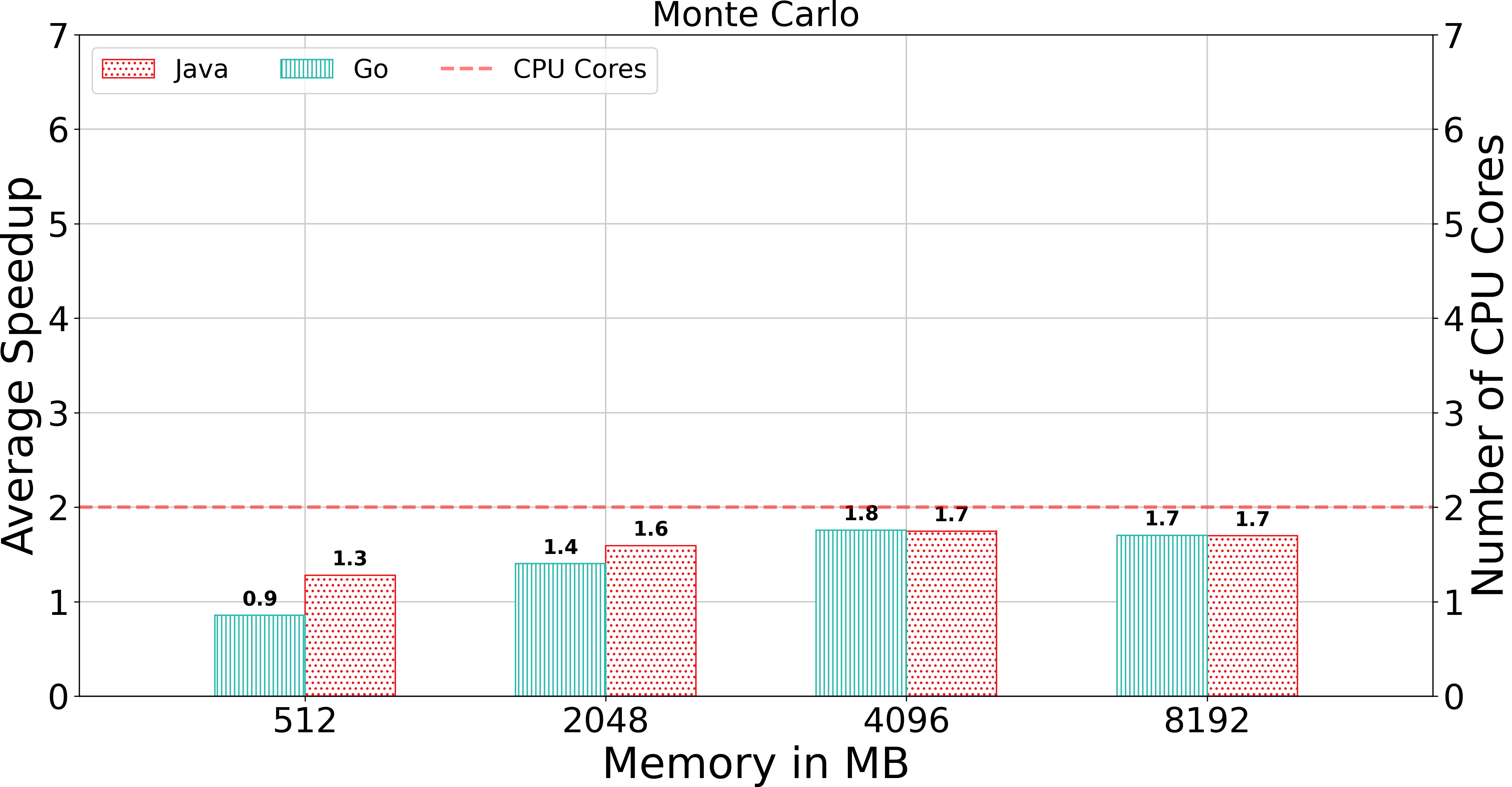}
        \caption{GCF Monte Carlo}
        \label{fig:speedups_gcf_monte}
    \end{subfigure}
    \begin{subfigure}{0.32\textwidth}
        \includegraphics[width=1\linewidth]{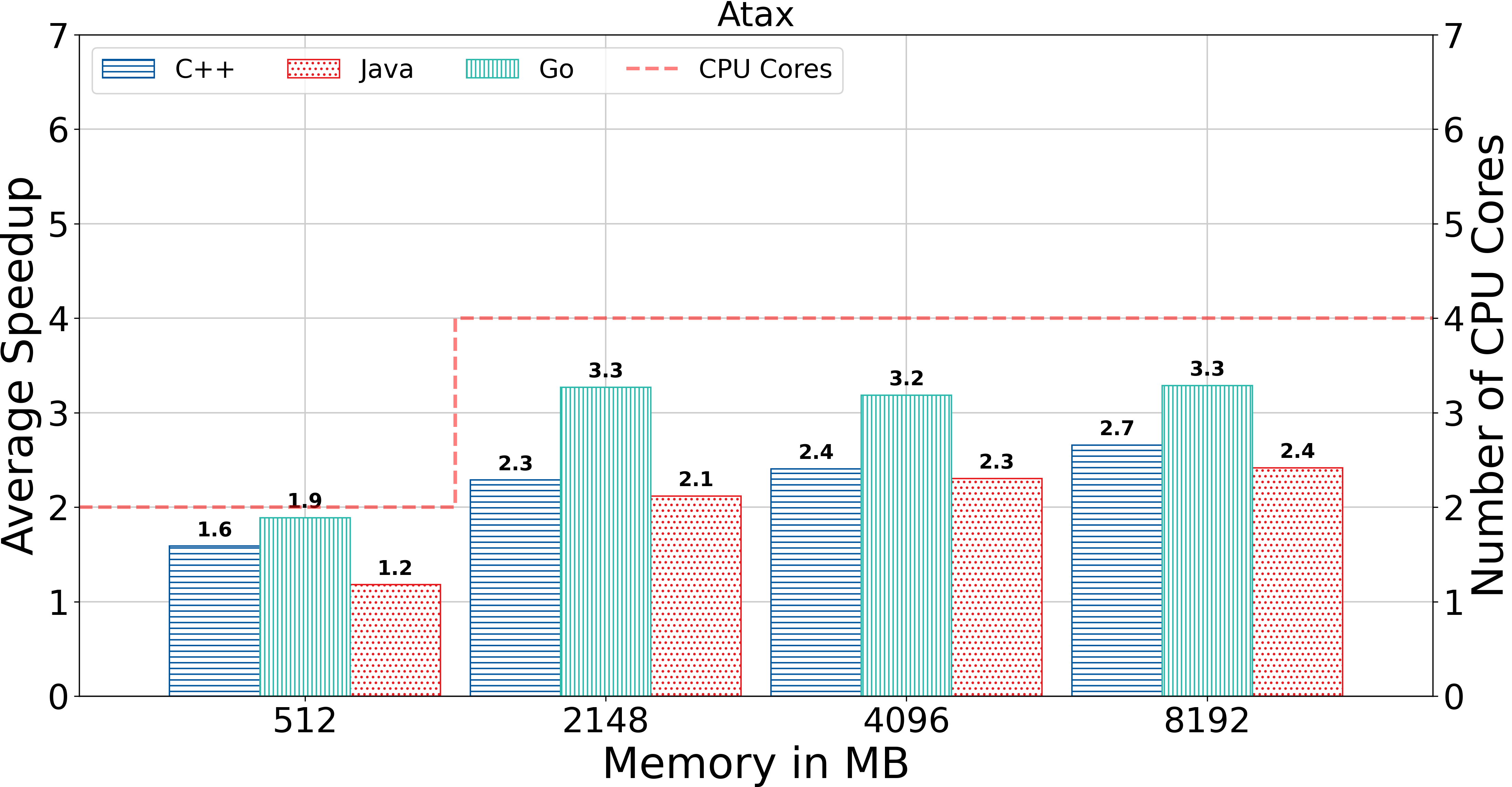}
        \caption{GCR Atax}
        \label{fig:speedups_gcr_atax}
    \end{subfigure}
    \begin{subfigure}{0.32\textwidth}
        \includegraphics[width=1\linewidth]{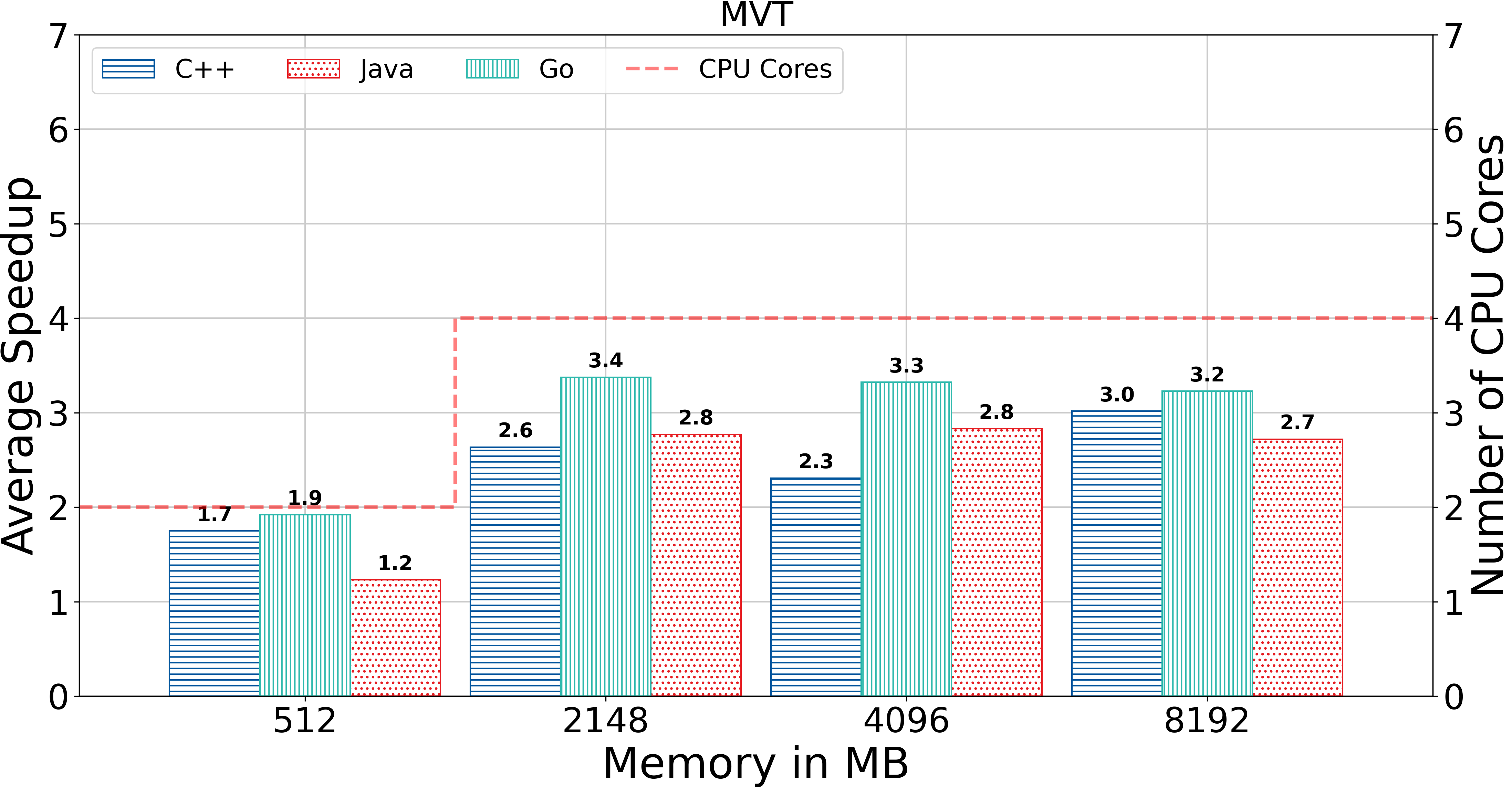}
        \caption{GCR MVT}
        \label{fig:speedups_gcr_mvt}
    \end{subfigure}
    \begin{subfigure}{0.32\textwidth}
        \includegraphics[width=1\linewidth]{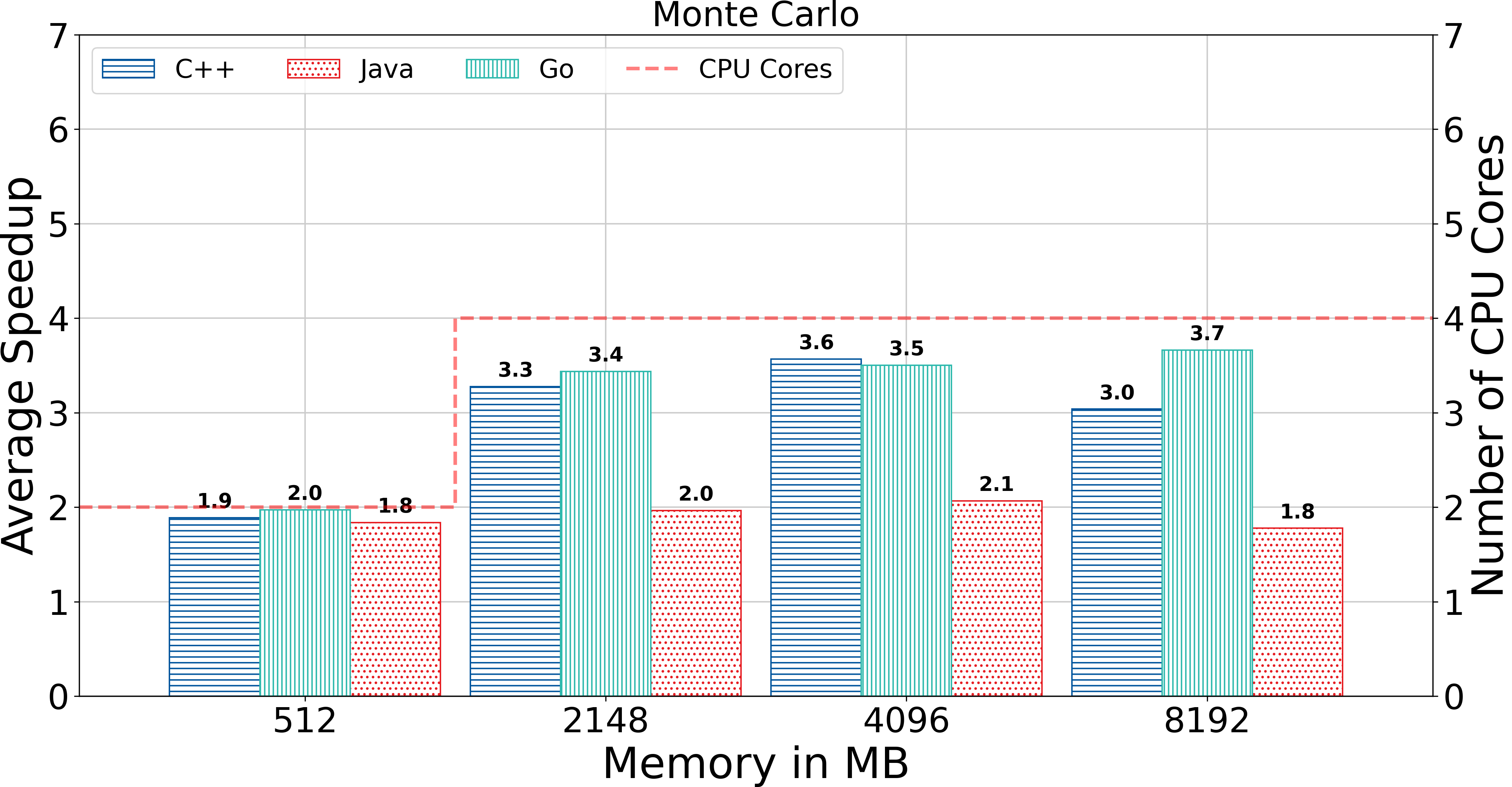}
        \caption{GCR Monte Carlo}
        \label{fig:speedups_gcr_monte}
    \end{subfigure}
    \caption{Obtained average speedups for the different parallelized workloads on AWS Lambda, GCF, and GCR. For a particular memory configuration, the red line shows the ideal speedup wrt the number of available CPU cores.}
    
    \label{fig:performance}
\end{figure*}

\subsection{Experimental Setup}
\label{sec:setup}


For GCF and Lambda, we deployed the different workloads (\S\ref{sec:faas_wokloads}) using the 
memory profiles $512$MB, $2048$MB, $4096$MB, and $8192$MB. For Lambda, we also utilized the highest available memory 
configuration for a function, i.e., $10240$MB. In contrast to GCF and Lambda, GCR allows developers to configure the number of vCPUs allocated to a function along with the memory. For GCR, we used similar memory configurations as GCF except for $2048$MB. In this case, we configured the workloads with $2148$MB of memory since that is the minimum memory required to allocate 4vCPUs to a function. We allocate 4vCPUs for every memory configuration in GCR except for $512$MB where we allocate 2vCPUs. To reduce variance in performance measurements for the serverless workloads due to cold starts~\cite{10.1145/3447545.3451173}, we set the maximum number of concurrent instances for all services to one. Furthermore, we set the maximum number of concurrent requests that can be handled by a container in GCR to one. This is done to prevent the sharing of vCPUs while handling multiple simultaneous requests. We deploy all 
functions on GCR and GCF in the \texttt{us-central1} region. For Lambda, all functions are deployed in the \texttt{us-east1} region.

\subsection{\#CPU cores to vCPU mapping}
\label{sec:mapping}

For the different services, language runtimes (\S\ref{sec:language_runtimes}), and configurations (\S\ref{sec:setup}), we identified the number of available CPU cores. We obtained the number of available cores for the function/container instance using 
the Linux \texttt{proc} filesystem. The number of available CPU cores for the different services at the different memory profiles is shown in Figure~\ref{fig:numbercores}. For Lambda, we observed at least two CPU cores for every memory configuration. Lambda allocates one full vCPU per $1769$MB of allocated function memory~\cite{LambdaConfig}. This implies that the amount of allocated vCPUs is not equal to the number of available CPU cores. For instance, for a memory configuration of $512$MB on Lambda, we observed two CPU cores while not getting a full allocated vCPU. Moreover, Lambda always rounds up the number of available CPU cores as shown in Figure~\ref{fig:numbercores}. For example, $4096$MB translates to 2.3vCPUs, but we observed three CPU cores for that specific configuration. For GCF, we always observed two CPU cores irrespective of the configured function memory. Similarly for GCR, we observed at least two CPU cores for the different configurations (\S\ref{sec:setup}). However, for the container with the Java language runtime (\S\ref{sec:language_runtimes}), we observed only one CPU core when configured with 1vCPU. For greater than 1vCPU allocations in GCR, the number of available CPU cores is always equal to the number of configured vCPUs. Note that although not shown in Figure~\ref{fig:numbercores}, AWS Lambda provides four CPU cores for function instances configured with $6$GB of memory.

A possible explanation for observing two CPU cores at lower memory configurations for the different services, i.e., AWS Lambda and GCF can be hyperthreading or Simultaneous Multithreading (SMT)~\cite{eggers1997simultaneous} present in modern Intel Server Family of Processors, i.e, Haswell-EP, Broadwell-EP, Skylake-SP, and Cascade Lake-SP. As shown in previous works~\cite{chadha2021architecture, behind}, these are the family of processors found in the Virtual Machines of the commercial FaaS providers on which the function/container instances are launched.

\begin{figure*}[t]
    \centering
    \begin{subfigure}{0.32\textwidth}
        \includegraphics[width=1\linewidth]{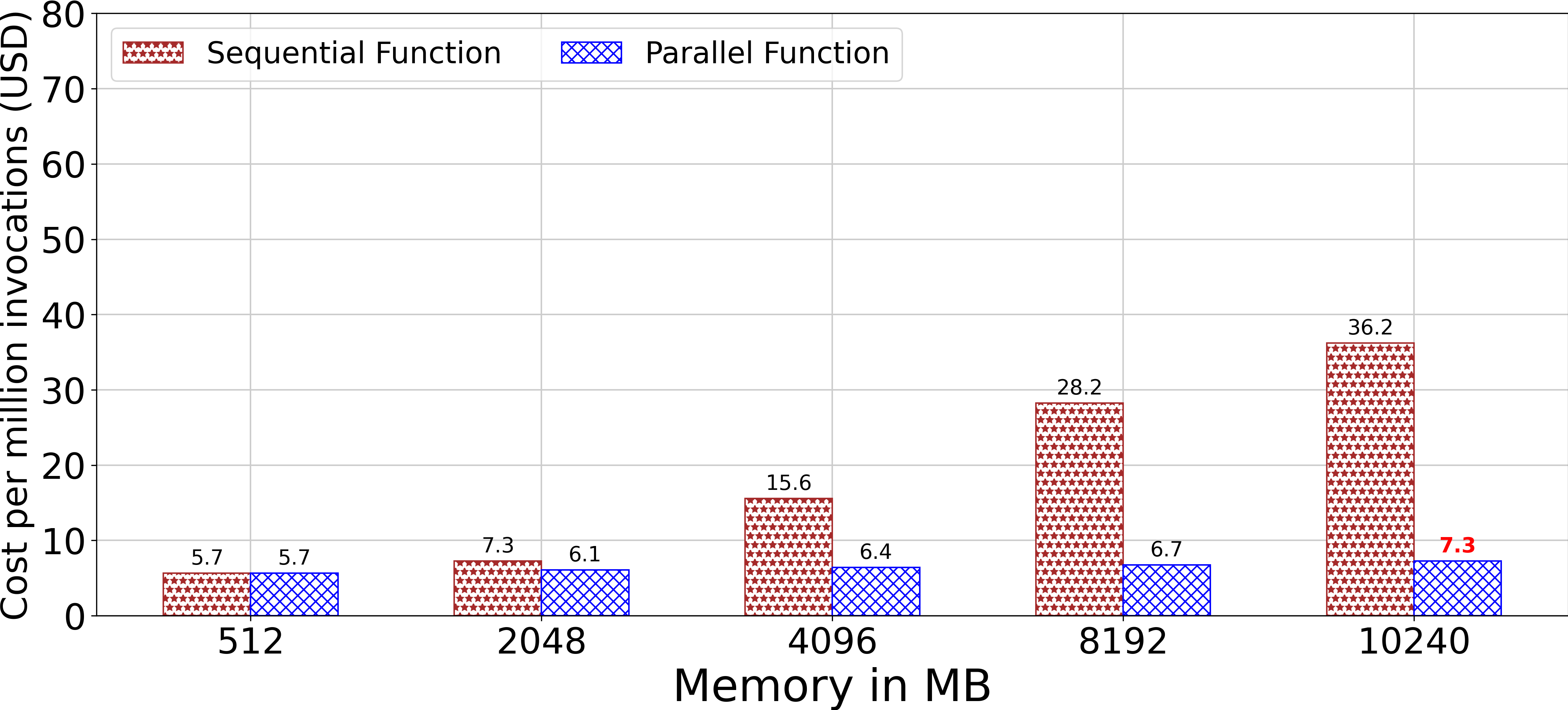}
        \caption{AWS Atax C++}
        \label{fig:cost_aws_atax_c++}
    \end{subfigure}
    \begin{subfigure}{0.32\textwidth}
        \includegraphics[width=1\linewidth]{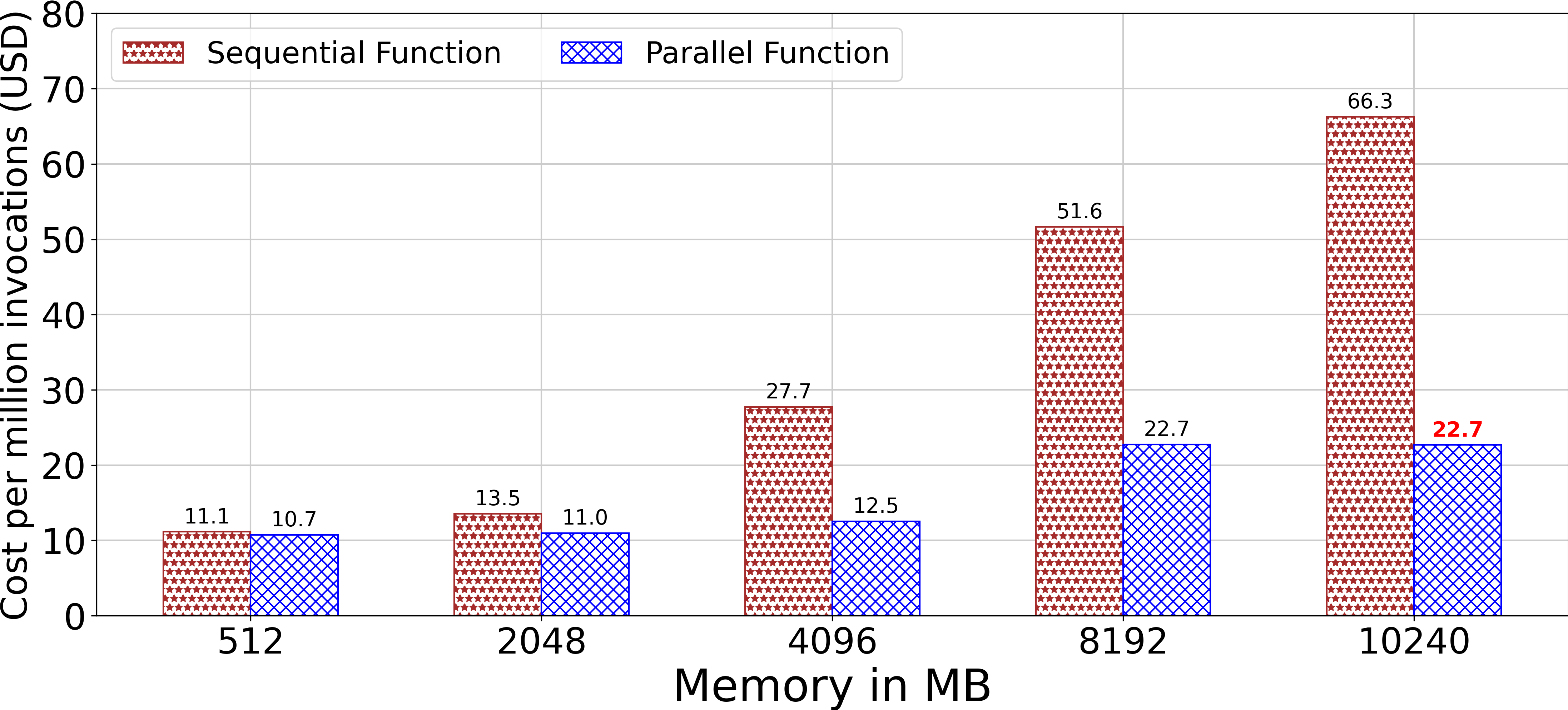}
        \caption{AWS Atax Go}
        \label{fig:cost_aws_atax_go}
    \end{subfigure}
    \begin{subfigure}{0.32\textwidth}
        \includegraphics[width=1\linewidth]{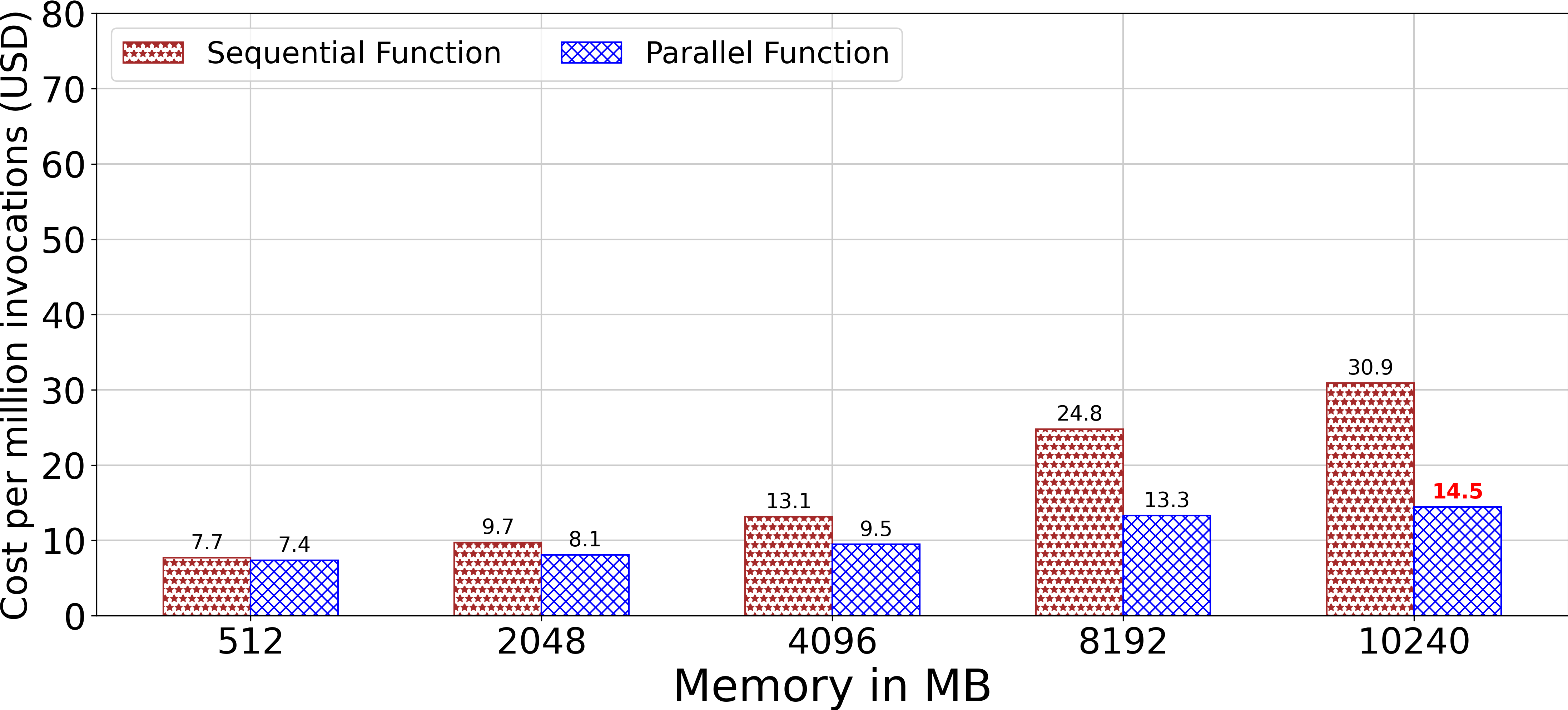}
        \caption{AWS Atax Java}
        \label{fig:cost_aws_atax_java}
    \end{subfigure}
    \begin{subfigure}{0.32\textwidth}
        \includegraphics[width=1\linewidth]{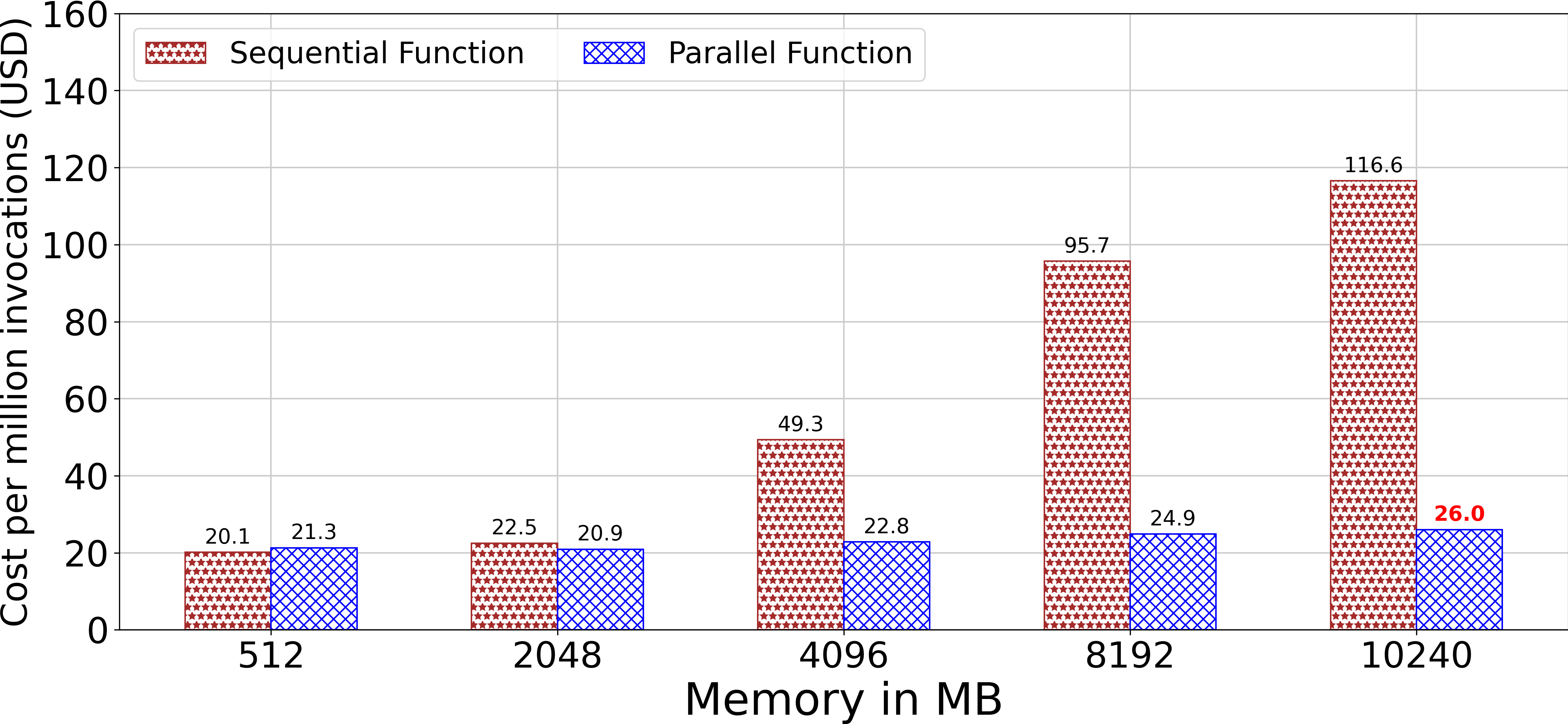}
        \caption{AWS MVT C++}
        \label{fig:cost_aws_mvt_c++}
    \end{subfigure}
    \begin{subfigure}{0.32\textwidth}
        \includegraphics[width=1\linewidth]{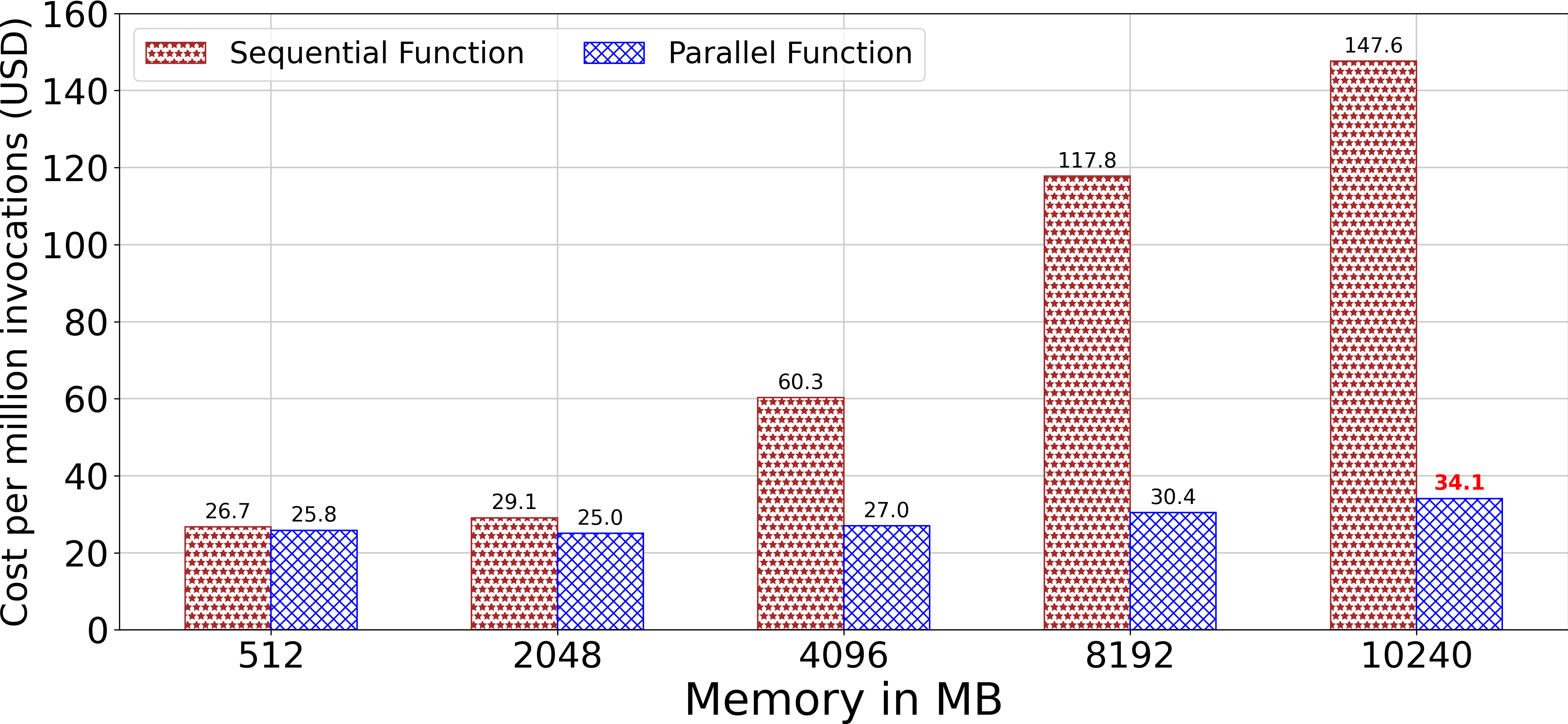}
        \caption{AWS MVT Go}
        \label{fig:cost_aws_mvt_go}
    \end{subfigure}
    \begin{subfigure}{0.32\textwidth}
        \includegraphics[width=1\linewidth]{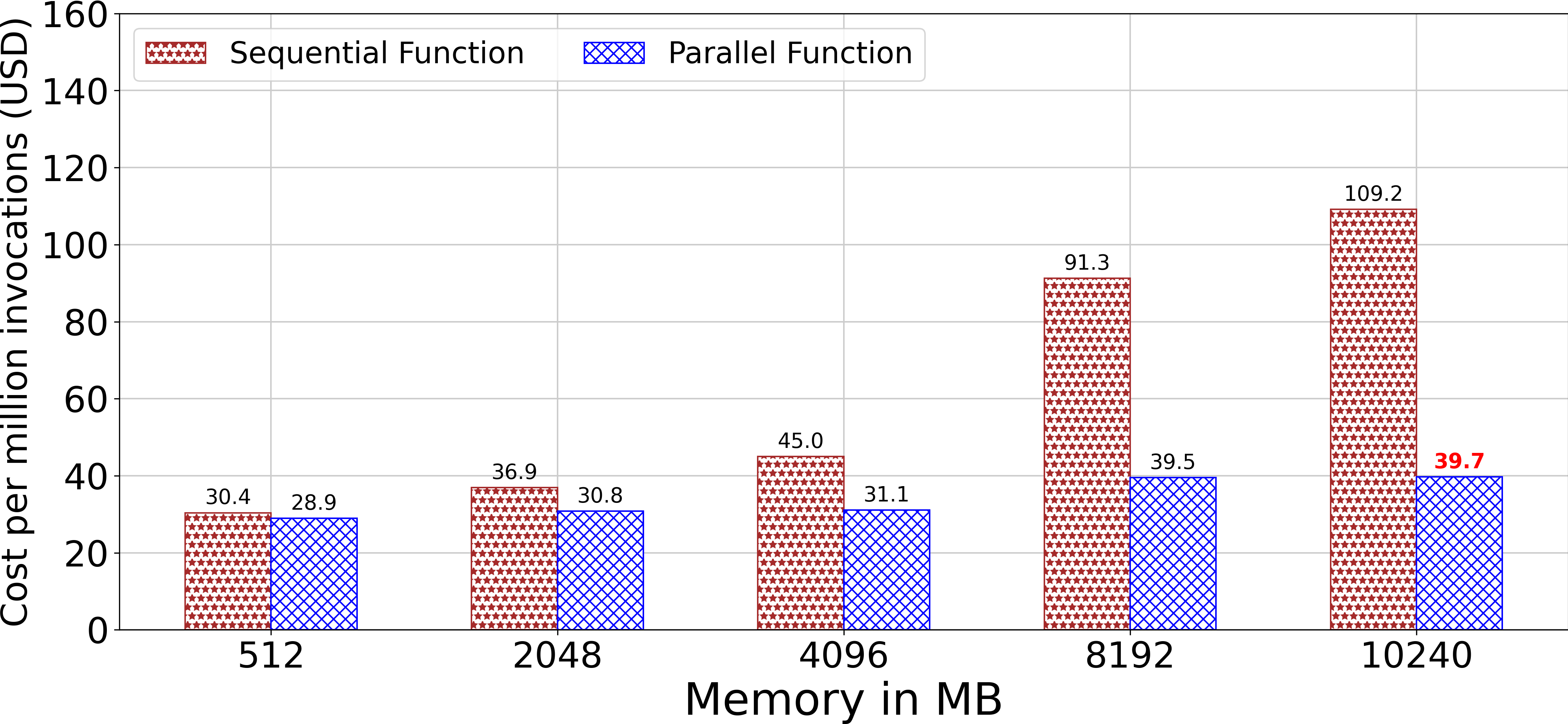}
        \caption{AWS MVT Java}
        \label{fig:cost_aws_mvt_java}
    \end{subfigure}
    \begin{subfigure}{0.32\textwidth}
        \includegraphics[width=1\linewidth]{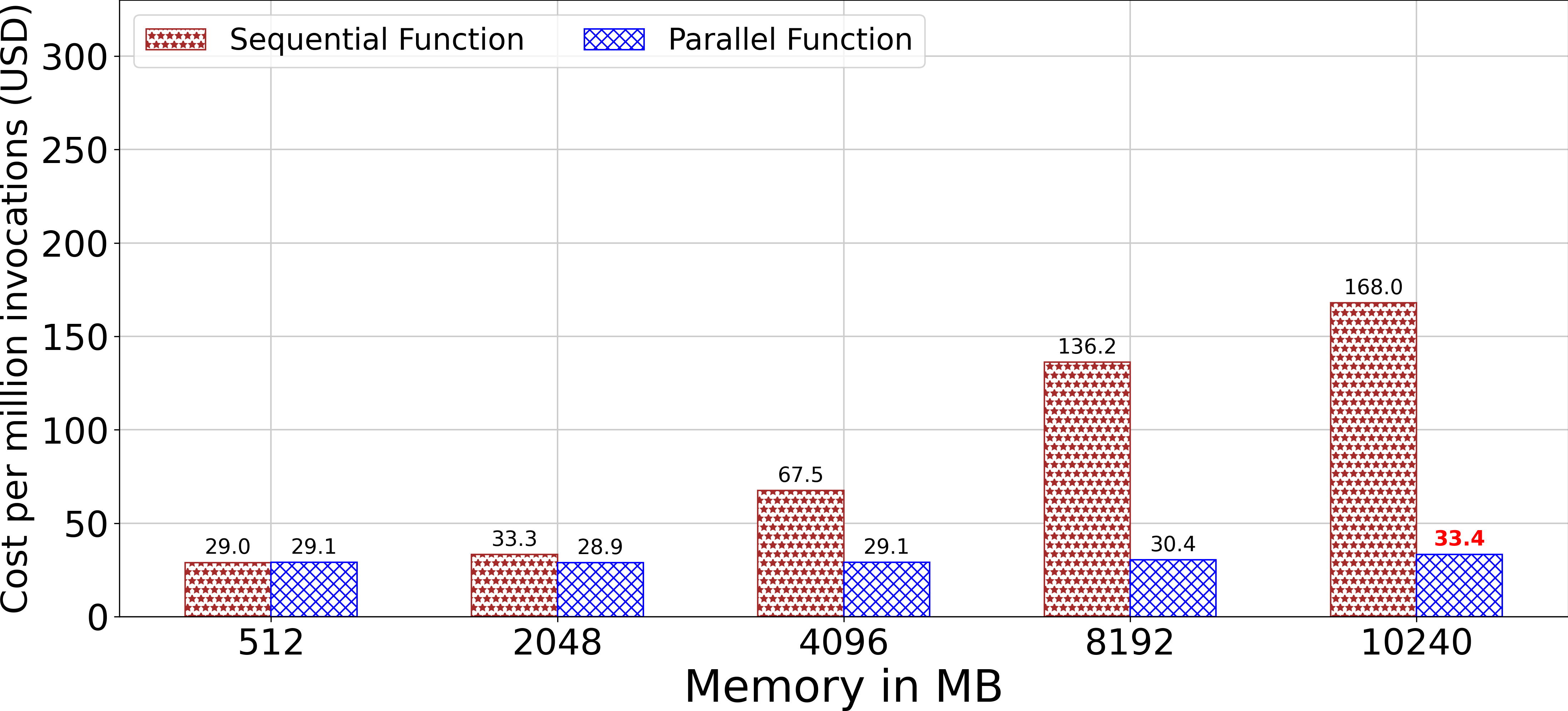}
        \caption{AWS Monte Carlo C++}
        \label{fig:cost_aws_monte_c++}
    \end{subfigure}
    \begin{subfigure}{0.32\textwidth}
        \includegraphics[width=1\linewidth]{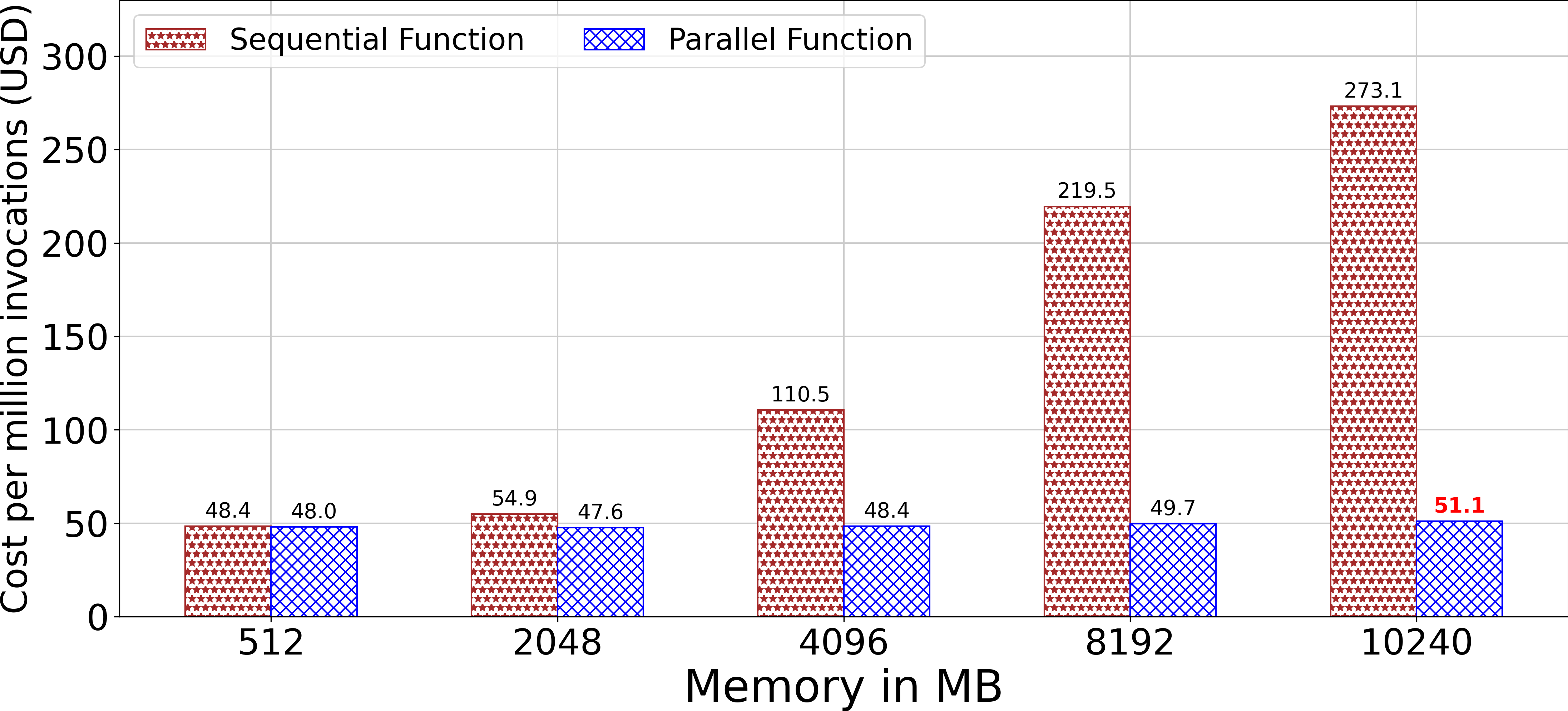}
        \caption{AWS Monte Carlo Go}
        \label{fig:cost_aws_monte_go}
    \end{subfigure}
    \begin{subfigure}{0.32\textwidth}
        \includegraphics[width=1\linewidth]{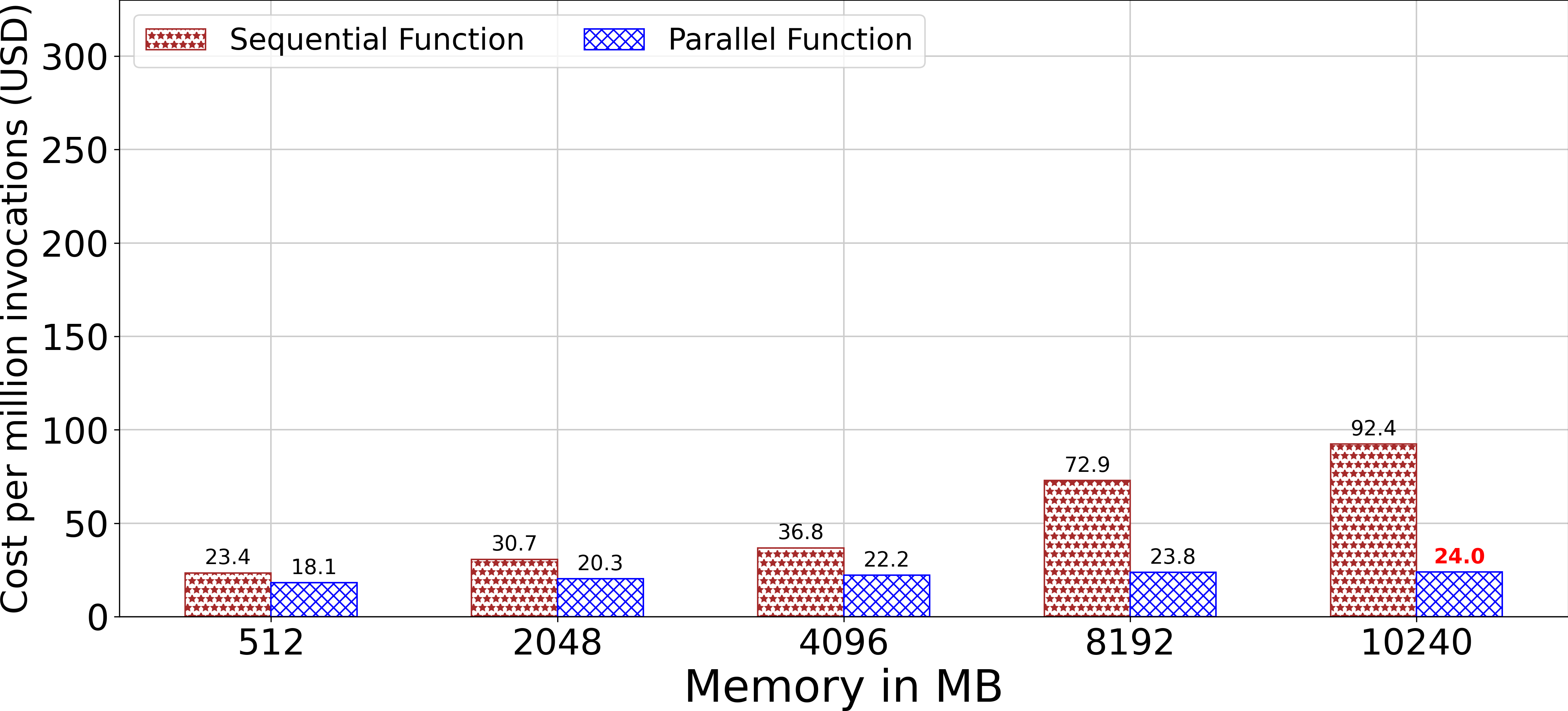}
        \caption{AWS Monte Carlo Java}
        \label{fig:cost_aws_monte_java}
    \end{subfigure}
    \caption{Comparison of cost per million function invocations (in USD) for the different workloads on Lambda. The cost values highlighted with red represent the maximum cost savings obtained across the different memory configurations.}
    \label{fig:costs}
\end{figure*}

\subsection{Comparing Performance}
\label{sec:comparing_perf}

Figure~\ref{fig:performance} shows the average speedups obtained for the different serverless workloads (\S\ref{sec:faas_wokloads}) across the different memory configurations (\S\ref{sec:setup}) for the different services. For a particular memory configuration, we compute the speedup obtained by dividing the mean execution time of the sequential serverless workload by the mean execution time of the parallelized version. Note that, when executing a parallelized workload, we use the maximum number of available CPU cores for a particular memory configuration (\S\ref{sec:mapping}). For Lambda and GCF, we don't observe any significant speedup for lower memory configurations, i.e., less than $2048$MB. This is because, in Lambda and GCF, each function instance has a fixed memory and fraction of allocated CPU cycles. Since both FaaS platforms do not allocate complete two vCPUs for lower memory configurations, utilizing the underlying two CPU cores does not improve performance. For Lambda, two vCPUs are allocated for a memory configuration greater than $3538$MB, while for GCF they are allocated at a memory configuration of $4096$MB~\cite{GCFPricing}. This is also apparent from the increase in speedup observed for GCF when switching from $2048$MB to $4096$MB as shown in Figures~\ref{fig:speedups_gcf_atax},~\ref{fig:speedups_gcf_mvt}, and~\ref{fig:speedups_gcf_monte}. For memory configurations greater than $4096$MB, we observed speedup close to the number of available vCPUs for both GCF and Lambda. This shows that irrespective of the number of available CPU cores to a function instance, the performance of a parallelized function is limited by the allocated vCPUs.

\begin{table}[t]
    \caption{Maximum obtained cost savings, language runtime, and memory configuration for the different serverless workloads on GCF/GCR.}
	\centering
    \begin{adjustbox}{width=8.5cm,center}
    \begin{tabu}{|c|c|c|c|c|}
	\tabucline{-}

    \textbf{Service} & \textbf{Benchmark} & \textbf{Max. Cost savings} & \textbf{Runtime} & \textbf{Memory} \\\tabucline{-}
    GCF & Atax & 49.1\% & Java & 4096MB \\ \tabucline{-}
    GCF & MVT & 39.7\% & Java & 4096MB \\ \tabucline{-}
    GCF & Monte Carlo & 44.1\% & Java & 8192MB \\ \tabucline{-}
    GCR & Atax & 59.5\% & Go & 2148MB \\ \tabucline{-}
    GCR & MVT & 63.4\% & Go & 2148MB \\ \tabucline{-}
    GCR & Monte Carlo & 69.8\% & Go & 2148MB \\ \tabucline{-}
  
\end{tabu}
\end{adjustbox}
\label{tab:costSavings}
\end{table}

For Lambda, the parallelized versions of \texttt{Atax} and \texttt{MVT} microbenchmarks perform consistently worse than \texttt{Monte Carlo} for higher memory configurations. This is because both these microbenchmarks have a greater number of parallel regions than \texttt{Monte Carlo} and require synchronization and communication between the application threads. Moreover, the Go and C++ implementations of the serverless workloads perform better than the Java based implementation as shown in Figures~\ref{fig:speedups_aws_atax},~\ref{fig:speedups_aws_mvt},~\ref{fig:speedups_aws_monte},~\ref{fig:speedups_gcr_atax},~\ref{fig:speedups_gcr_mvt}, and~\ref{fig:speedups_gcr_monte} . This can be attributed to the performance degradation of parallel implementations using Java threads with an increase in the number of application threads and communication~\cite{10.1145/1596655.1596661}. For GCF, the Go implementations perform worse as compared to Lambda and GCR, since GCF uses an old runtime version for Go. As a sanity check, we could reproduce the results by using \texttt{golang:1.13-buster} as the container image for Go with GCR. For GCR, speedup values obtained are similar to that for Lambda, i.e., close to the number of allocated vCPUs and therefore capped at four. Note that for GCR, we even obtained speedup values for the lowest memory configurations since we were able to allocate two complete vCPUs. The observed performance and speedup depend on the parallelization efficiency of the different serverless workloads~(\S\ref{sec:faas_wokloads}). 





\subsection{Comparing Costs}

To calculate costs for the different serverless workloads (\S\ref{sec:faas_wokloads}), we use the obtained mean execution time across the different memory configurations. For Lambda, we round up the execution time to the nearest 1ms increment, while for GCF and GCR it is rounded up to the nearest 100ms increment. Following this, we use the rounded up mean execution time to calculate function compute time in terms of different units defined by the providers~\cite{LambdaPricing, GCFPricing, GCRPricing}. For Lambda, the compute time depends on the amount of allocated memory, i.e., GB-Seconds, for GCF it depends on the configured memory and the allocated CPU clock cycles, i.e., GB-Seconds and GHz-Seconds, and for GCR it depends on the configured memory and the allocated vCPUs, i.e, GB-Seconds and vCPU-Seconds. The different providers define a fixed price for one second of compute time depending on the deployment region. We use the pre-defined price values specified by all cloud providers for calculating the compute costs for the serverless workloads. In our calculations, we exclude costs for free-tiers and networking. Moreover, we calculate costs per million function invocations. As a result, a fixed price of \$$0.2$ and \$$0.4$ is added for Lambda and GCF/GCR respectively. 

Figure~\ref{fig:costs} shows the cost comparison for the sequential and parallelized serverless workloads (\S\ref{sec:faas_wokloads}) across the different memory configurations on Lambda. For Lambda, the difference in costs for the sequential and parallelized serverless workloads is not significant for lower memory configurations, i.e., less than $2048$MB. However, for the sequential serverless workloads the costs significantly increase for higher memory configurations, while for the parallelized versions the increase in cost is considerably less. For instance, the increase in average cost from the lowest to the highest memory configuration for the workloads parallelized using C++ is $30$\%, while for the sequential versions it is $498$\%. Therefore, by efficiently parallelizing serverless workloads we can obtain improved performance (\S\ref{sec:comparing_perf}) at approximately the same costs. Overall, we observed that parallelizing functions with C++ leads to minimum costs as shown in Figure~\ref{fig:costs}. For the sequential workloads, Java is the cheapest, except for lower memory configurations. For Lambda, we obtained maximum cost savings of $81$\% for the Go implementation of the \texttt{Monte Carlo} workload as shown in Figure~\ref{fig:cost_aws_monte_go}.

Due to space limitations, we do not present detailed cost analysis results for GCF/GCR but summarize our findings in Table~\ref{tab:costSavings}. We obtain the maximum cost savings for GCF with a memory allocation of at least $4096$MB which also corresponds to the highest obtained speedup values (\S\ref{sec:comparing_perf}). For GCR, the maximum cost savings are obtained for the memory configuration of $2148$MB with $4$vCPUs. 


\subsection{Impact of Cold Starts}
\label{sec:impact_cold_starts}

\begin{figure}[t]
    \centering
    \includegraphics[width=1\columnwidth]{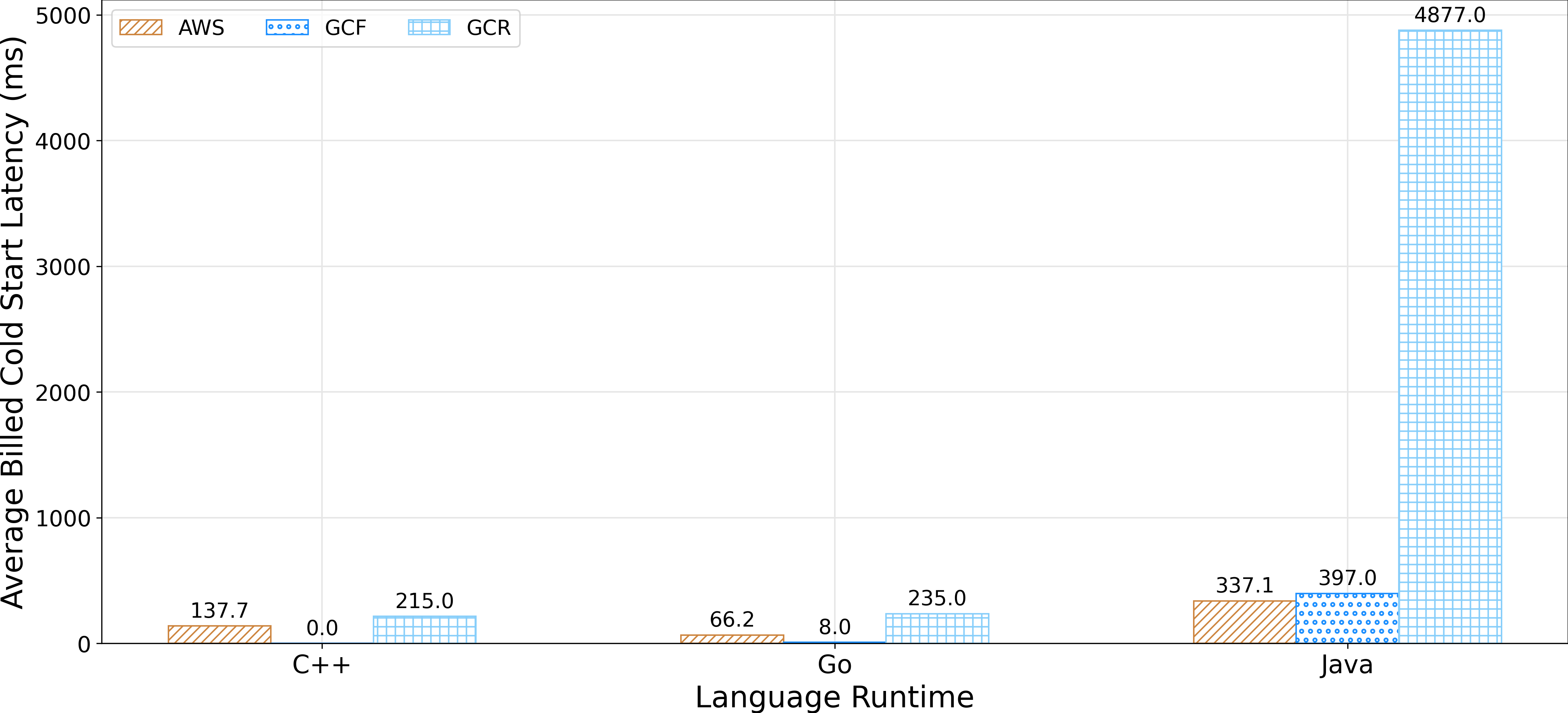}
    \caption{Average billable cold start latency for the different runtimes across the different services.}
    \label{fig:coldstart}
\end{figure}

From our experiments, we observed that the latency of cold starts is equally long for the sequential and parallel implementations of the serverless workloads (\S\ref{sec:faas_wokloads}). Since cold starts constitute a fraction of the function execution time, if large enough, they can potentially impact the speedup and cost savings obtained from parallelizing serverless workloads.  The average billable cold start latency for the different services and language runtimes (\S\ref{sec:language_runtimes}) is shown in Figure~\ref{fig:coldstart}. While AWS Lambda directly provides the billed cold start time in function logs, for GCF and GCR, we compute the billed cold start latency by subtracting the billed function execution time on a cold and warm start. In our experiments, we did not observe a significant difference in the cold start latencies for the different serverless workloads and function memory configurations. However, the values shown in Figure~\ref{fig:coldstart} are averaged across all benchmarks. We observed a significant cold start latency for the Java runtime on GCR as compared to the C++ and Go runtimes. A possible explanation for this could be a high amount of time required for the Java Virtual Machine (JVM) to warm up.

\section{Conclusion \& Future Work}
\label{sec:conclusion}

In this paper, we analyzed the effect of parallelizing compute-intensive serverless workloads within a function/container instance in terms of performance and costs for AWS Lambda, Google Cloud Functions, and Google Cloud Run. We identified that for the different services the number of CPU cores available to the function/container does not always equal the number of allocated vCPUs. Furthermore, we demonstrate that parallelizing serverless workloads can significantly improve performance and lead to cost savings. For Lambda, we observed cost savings up to 81\%, for GCF up to 49\%, and for GCR up to 79.8\%. 

In the future, we plan to evaluate other serverless offerings and language runtimes. Furthermore, investigating a hybrid approach between inter and intra-function parallelism to reduce synchronization overhead is another future direction.

\section{Acknowledgement and Reproducibility}
This work was supported by the funding of the German Federal Ministry of Education and Research (BMBF) in the scope of the Software Campus program. 

All code artifacts related to this work are available at\footnote{https://github.com/MichaelKiener/Serverless-parallel-workloads}. We also evaluated other parallelized workloads which could not be included in this paper due to page limitations. Please refer to~\cite{michaelkienerthesis} for our comprehensive evaluation results.

\bibliographystyle{ACM-Reference-Format}
\bibliography{serverless}


\end{document}